\documentclass[aps,tightenlines,twocolumn,superscriptaddress,sort&compress]{revtex4}

\usepackage{amssymb}
\usepackage{color,graphicx}
\usepackage{amsmath}
\usepackage{mathtools}
\usepackage{amsbsy}
\usepackage{bbm}
\usepackage{bm}
\usepackage{comment}
\usepackage{booktabs}
\usepackage{array}

\newcommand{\beq}{\begin{eqnarray}}
\newcommand{\eeq}{\end{eqnarray}}

\newcommand{\ve}{\mathbf}
\DeclarePairedDelimiter{\abs}{\lvert}{\rvert}
\DeclarePairedDelimiter{\norma}{\lVert}{\rVert}
\begin{document}

\title{Bounds on collapse models from cold-atom experiments}

\author{M. Bilardello}
\email{marco.bilardello@ts.infn.it}
\affiliation{Department of Physics, University of Trieste, Strada Costiera 11, 34014 Trieste, Italy}
\affiliation{Istituto
Nazionale di Fisica Nucleare, Trieste Section, Via Valerio 2, 34127 Trieste,
Italy}

\author{S. Donadi}
\email{sandro.donadi@ts.infn.it}
\affiliation{Department of Physics, University of Trieste, Strada Costiera 11, 34014 Trieste, Italy}
\affiliation{Istituto
Nazionale di Fisica Nucleare, Trieste Section, Via Valerio 2, 34127 Trieste,
Italy}

\author{A. Vinante}
\email{anvinante@fbk.eu}
\affiliation{Istituto di Fotonica e Nanotecnologie, CNR - Fondazione Bruno Kessler, I-38123 Povo, Trento, Italy}

\author{A. Bassi}
\email{bassi@ts.infn.it}
\affiliation{Department of Physics, University of Trieste, Strada Costiera 11, 34014 Trieste, Italy}
\affiliation{Istituto
Nazionale di Fisica Nucleare, Trieste Section, Via Valerio 2, 34127 Trieste,
Italy}

\begin{abstract}
The spontaneous localization mechanism of collapse models induces a Brownian motion in all physical systems. This effect is very weak, but  experimental progress in creating ultracold atomic systems can be used to detect it. In this paper, we considered a recent experiment~\cite{kas}, where an atomic ensemble was cooled down to picokelvins. Any Brownian motion induces an extra increase of the position variance of the gas. We study this effect by solving the dynamical equations for the Continuous Spontaneous Localizations (CSL) model, as well as for its non-Markovian and dissipative extensions. The resulting bounds, with a $95 \%$ of confidence level,  are beaten only by measurements of spontaneous X-ray emission and by experiments with cantilever (in the latter case, only for $r_C \geq 10^{-7}$ m, where $r_C$ is one of the two collapse parameters of the CSL model). We show that, contrary to the bounds given by X-ray measurements, non-Markovian effects do not change the bounds, for any reasonable choice of a frequency cutoff in the spectrum of the collapse noise. Therefore the bounds here considered are more robust.  We also show that dissipative effects are unimportant for a large spectrum of temperatures of the noise, while for low temperatures the excluded region in the parameter space is the more reduced, the lower the temperature.    
\end{abstract}

%\pacs{03.67.Mn}

\maketitle

%%%%%%%%%%%%%%%%%%%%%%%%%%%%%%%%%%%%%%%%%%%%%%%%%%%%%%%%%%%%%%%%%%%%%
\section{Introduction}
%%%%%%%%%%%%%%%%%%%%%%%%%%%%%%%%%%%%%%%%%%%%%%%%%%%%%%%%%%%%%%%%%%%%%%
The accuracy of experiments testing the quantum properties of larger and larger systems is improving at a fast pace. Quantum superpositions have been directly  observed in a large variety of mesoscopic systems, ranging from atoms~\cite{anderson,andrews} to macro-molecules~\cite{9999, 10000,romero}, and optomechanics promises to reach much larger masses~\cite{osc1,osc2,osc3}. This is interesting and important, since it helps answering the question whether the quantum superposition principle, the building block of the theory, holds also at large scales, or breaks down at some point. Moreover, as most quantum technologies rely on the superposition principle being applicable to arbitrarily complex systems, assessing its validity will impact the future directions of technological research.

On the theoretical side, collapse models~\cite{grw, pearleold, qmupl, rep1, rep2, sci} take into account, in a quantitative way, the possibility of a progressive breakdown of  quantum linearity when the size and complexity of the system increase. More than this, strong arguments~\cite{gis, Adlerbook} show that they are the only possible way of modifying quantum theory, taking into account such a breakdown. Therefore, testing these models serve as a benchmark for any test of  the superposition principle. 

According to collapse models, material particles interact with an external classical noise, which induces the collapse of the wave function. The effect  is negligible for microscopic systems but, as it scales with the number of particles, macroscopic objects are always well-localised in space. The most complete and well studied  model is the Continuous Spontaneous Localization (CSL) model~\cite{csl}, which we will consider in this article together with its non-Markovian (cCSL)~\cite{nonwhite1, nonwhite2} and dissipative (dCSL)~\cite{andrea} extensions. 

The CSL model contains two new parameters: $\lambda$, which sets the strength of the interaction with the collapse noise, and $r_C$, which defines the resolution of the collapse process. As for the dCSL model, a third parameter $\kappa$, related to the temperature $T_{\text{\tiny{CSL}}}$ of the collapse noise, is introduced. In the cCSL model instead, the third new parameter is the cut-off frequency $\Omega$, which controls the noise spectrum. Setting a bound on $\lambda$ and $r_C$ (and on $T_{\text{\tiny{CSL}}}$ for dCSL model, and $\Omega$ for the cCSL model) is one of the outputs of experimental tests of the quantum superposition principle. 

In the literature, the following values for $\lambda$ and $r_C$ have been suggested. According to Ghirardi, Rimini and Weber~\cite{grw}, $r_C=10^{-7}$m and $\lambda \simeq 10^{-16}$ s$^{-1}$. These values come from the  requirement that macroscopic objects must alway be well localized. In~\cite{csl}, slightly different values were proposed:  $r_C=10^{-7}$m and $\lambda \simeq 10^{-17}$ s$^{-1}$. Adler, on the other hand, suggested stronger values:  $r_C=10^{-7}$m and $\lambda \simeq 10^{-8 \pm 2}$ s$^{-1}$ and  $r_C=10^{-6}$m and $\lambda \simeq 10^{-6 \pm 2}$ s$^{-1}$, as a result of the analysis of the process of latent image formation in photography~\cite{ad1}. With reference to the dCSL model, if the collapse noise is associated to some 
cosmological field, a reasonable value of the temperature of the noise field is $T_{\text{\tiny{CSL}}} \simeq 1$ K. For the cCSL model, a reasonable cosmological value for the frequency cut-off is $\Omega \simeq 10^{10} - 10^{11}$ Hz~\cite{bas}.

Upper bounds on the collapse parameters are set by experiments. The direct way of testing them is through interferometric experiments. The best limits of this kind come from matter-wave interferometry performed by Arndt's group~\cite{arbest}, which are reported in Fig.~\ref{exc_csl}. More recently, non interferometric experiments have been  pushed forward~\cite{bah, klaus, diosi2}. They all aim at testing a side-effect of the collapse noise: the Brownian motion it induces on the dynamics of any system.  Two experimental scenarios of this kind are relevant: cantilevers~\cite{vin}, where the Brownian motion shows up as a violation of the equipartition theorem (an anomalous heating), and X-ray detection~\cite{bea}, where the Brownian motion induces spontaneous photon emission from matter.  The relevant bounds are again reported in Fig.~\ref{exc_csl}.

A recent experiment~\cite{kas} succeeded in cooling a cloud of  $^{87}$Rb atoms down to pK. This serves as a further test of collapse models, as we will see. The authors of~\cite{kas} analyzed the spontaneous heating induced by a classical stochastic force acting on the cloud~\cite{kasapp}, and set a bound on the heating rate, due to the stochastic diffusion, equal to $20 \pm 30$ pK/s.

Aim of this article is to perform an exact calculation of the predictions of the CSL model for the experiment considered in~\cite{kas}, and compare these predictions with the experimental data. We will set bounds on $\lambda$ and $r_C$  of CSL (as well as on $T_{\text{\tiny{CSL}}}$  of dCSL and $\Omega$ of cCSL). In the case of dCSL we will see that there exist values of  $T_{\text{\tiny{CSL}}}$ such that the noise cools the system, not heat it. 

Instead of computing the change in the energy due to the collapse noise as done in~\cite{kas}, we will compute the change of the variance in position of the cloud, which is the quantity measured in the experiment.  The associated bounds are reported in Figs. 6 and 7. As we will see, these bounds are the strongest in a significant region of the parameter space, as we will discuss in Section IV. 

The paper is organized as follows. In section II we describe the experimental setup of~\cite{kas}. In section III we compute the theoretical predictions according to the CSL model. In sections IV and V we study, respectively, the predictions of the non-white and of the dissipative extensions of the CSL model. Finally, in section VI we compare the theoretical predictions with the experimental results, and we derive the upper bounds on the collapse parameters.

\section{Description of the experiment}
\begin{figure}[t]
\includegraphics[scale=0.4]{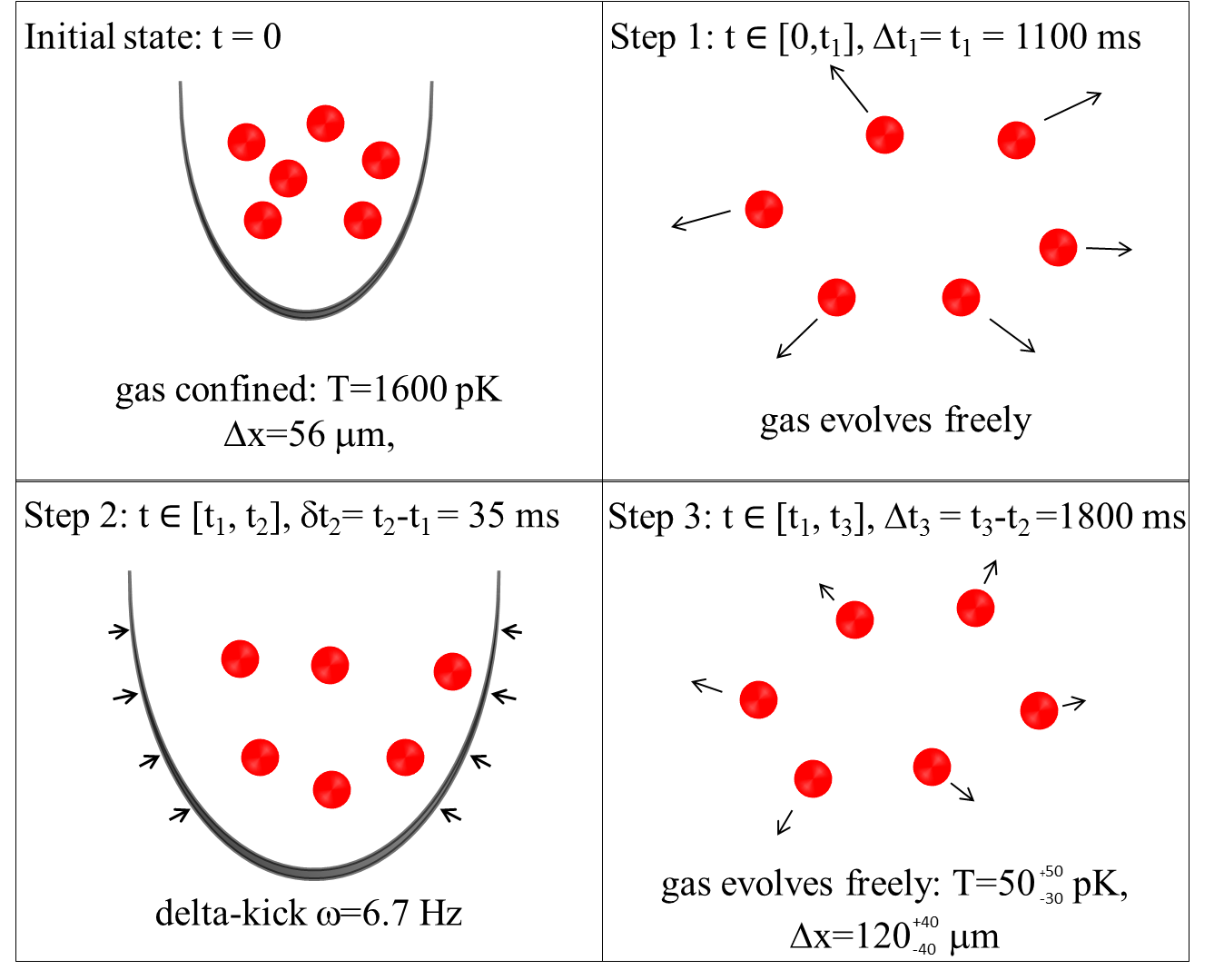}
\caption{Graphical representation of the experiment reported in~\cite{kas}. For each step, the relevant experimental data are given.}
\label{prima}
\end{figure}

%%%%%%%%%%%%%%%%%%%%%%%%%%%%%%%%%%%%%%%%%%%%%%%%%%%%%%%%%%%%%%%%%%%%%%
A gas of $^{87}$Rb atoms is cooled down to very low temperatures ($T =50_{-30}^{+50}$ pK) by using a ``delta-kick" technique. All the relevant experimental data are summarized in Fig.~\ref{prima}. The gas is initially (t=0) trapped by a harmonic potential with  standard deviation in position equal to $56\;\mu$m. The cooling process comprises the following three steps:

\noindent {\it Step 1}: The harmonic trap is removed and the gas evolves freely for a relatively long time, $\Delta t_1 = 1.1 $ s. This allows atoms with the same average momentum to be approximatively at the same distance from the initial localized state of the gas.

\noindent {\it Step 2}: Delta-kick. A Gaussian laser beam interacts with the atoms, the laser-atom interaction being modeled by an external harmonic  potential. By choosing the proper harmonic frequency and interaction time, the  potential reduces the kinetic energy of the atoms. The interaction lasts for a short time, $\delta t_2 \simeq 35$ ms.

\noindent  {\it Step 3}: The gas evolves again freely  for a relatively long time, $\Delta t_3= 1.8 $ s. The position variance of the gas is then measured, from which the temperature of the gas is inferred. 
%
\begin{comment}
\begin{table}[tb]
\footnotesize
\caption{Experimental values of the work described in~\cite{kas}. $T_{\text{in}}$ is the initial temperature, $T_{\text{fin}}$ is the minimum temperature achieved. $ \Delta \hat{\ve{x}}_t:=\sqrt{\langle \hat{\ve{x}}^2 \rangle_t - \langle \hat{\ve{x}} \rangle_{t}^{2}}$ is the standard deviation at time $t$. The values of $\tau_p$ and $\Delta \hat{\ve{x}}_{t_2}$ are those which minimize the final standard deviation.}
\label{experiment}
\centering
\vspace{0.3cm}
\begin{tabular}{ *{9}{|c} }
\toprule
\hline 
{$\Delta \hat{\ve{x}}_{0}$($\mu$m)} &  {$\Delta \hat{\ve{x}}_{t_2}$ ($\mu$m)} &{$T_{\text{in}}$ (pK)} & {$T_{\text{fin}}$(pK)} & {$t_1$(ms)} &{$\tau_p$(ms)} &{$t_2$(ms)} \\
\midrule
\hline
$56$ & $120 \pm 40$ & $ 1600 $& $ 50_{-30}^{+50}$ & $1100$ & $35 $ & $1800$  \\
\hline
\bottomrule
\end{tabular}
\end{table}
\end{comment}

The delta-kick frequency $\omega$ plays a critical role in the analysis. An estimation of $\omega$ is given in Eq.~(103) of~\cite{kasapp} through a classical calculation. If the initial position and velocity of the atoms are uncorrelated, the frequency becomes:
\begin{equation}\label{omkas}
\omega=\sqrt{\frac{1}{\delta t_{\text{\tiny{min}}}}\left(\frac{1}{\Delta t_3}+(1-\gamma^2)\frac{1}{\Delta t_1}\right)}
\end{equation}
with $\gamma^2=0.017$ and $\delta t_{\text{\tiny{min}}}\approx 34$ ms is the time when the gas reaches the minimum spread in position (obtained in~\cite{kas} through a fit of the experimental data). Inserting all numerical values, we obtain: $\omega \approx 6.53$ rad/s. 

As a confirmation of this prediction, we verified that, for all the values of $\omega$ outside the range 6-7 rad/s, the predicted increase of the variance $\langle\hat{\ve{x}}^2 \rangle_{ t_3}$ is in contradiction with the experimental data even for $\lambda=0$, i.e. even for ordinary quantum mechanics. Therefore, $\omega$ should lay within that interval. Then, we divided the interval 6-7 rad/s in ten parts, and computed  which of the ten values of $\omega$ gives the weakest bounds on $\lambda$ and $r_C$; the result is $\omega=6.7$ rad/s. Since we can not estimate the error associated to $\omega$, we take a conservative attitude, and choose  this value for the following calculations.

In Fig.~$3$ of~\cite{kas}, the experimental data are shown. However, the only experimental value, explicitly reported together with error-bars, is the minimum value of the position standard deviation, $120^{+40}_{-40} \, \mu$m,  detected at delta-kick time of $\delta t_2 = 35$ ms, and shown in the inset of our Fig.~\ref{pos-taup}. This is the experimental value we will use in section VI to compute the bounds on the collapse parameters.

% We will use this effect to compute the upper bounds on the collapse parameters. 

%%%%%%%%%%%%%%%%%%%%%%%%%%%%%%%%%%%%%%%%%%%%%%%%%%%%%%%%%%%%%%%%%%%%%
\section{Expansion of the gas according to the CSL  model}
%%%%%%%%%%%%%%%%%%%%%%%%%%%%%%%%%%%%%%%%%%%%%%%%%%%%%%%%%%%%%%%%%%%%%%
%
We compute the time evolution of the variance in position, as well as the increase of energy of the gas, during the cooling process described in the previous section, according to the CSL model. The effect of CSL  is to increase the temperature of the gas, and consequently its spread in position. 
  
The master equation of the CSL model has the well-known Lindblad form~\cite{Lindblad, GKS, Breuer,csl}: 
\begin{eqnarray}\label{cslmasteq}
\frac{d\hat{\rho}(t)}{dt} &=& -\frac{i}{\hbar} \left [\hat{H},\hat{\rho} (t) \right]+ \int d\ve{y} \, \Bigl[ \mathbb{L} (\ve{y}) \hat{\rho} (t) \mathbb{L}^{\dagger} (\ve{y}) \nonumber\\
\nonumber\\
&-&\frac{1}{2} \left \{\mathbb{L}^{\dagger} (\ve{y})\mathbb{L} (\ve{y}),\hat{\rho} (t) \right\} \Bigr].
\end{eqnarray}

where the Hamiltonian is:  
\begin{equation}
\label{harmonic}
\hat{H} =\sum_{\alpha = 1}^{N} \, \hat{H}_{\alpha}:= \sum_{\alpha= 1}^{N} \left(\frac{\hat{\ve{p}}_\alpha}{2m} +  \frac{1}{2} m \omega^2 \hat{\ve{x}}_{\alpha}^{2}\right).
\end{equation} 
and the Lindblad operators $\mathbb{L} (\ve{y})$, for an $N$-atom system, are~\cite{csl, pearle}
\begin{eqnarray}\label{lcsl}
\mathbb{L}(\ve{y}) &=& \sum_{\alpha= 1}^{N} \hat{L}_{\alpha} (\ve{y}):= \sqrt{\frac{\lambda A^{2}}{\pi^{3/2}r_{C}^{3}}}\sum_{\alpha= 1}^{N} e^{-\frac{\left \lvert \hat{\ve{x}}_{\alpha} - \ve{y} \right\rvert^2}{2 r_{C}^{2}}},\\
\nonumber\\
&=&\frac{\sqrt{\lambda A^{2}8\pi^{3/2}r_{C}^{3}}}{(2\pi\hbar)^{3}}\sum_{\alpha=1}^{N}\int d\ve{Q}\,e^{\frac{i}{\hbar}\ve{Q}\cdot(\hat{\ve{x}}_{\alpha}-\ve{y})}\, e^{-\frac{r_{C}^{2}}{2 \hbar^{2}} \ve{Q}^2 },\nonumber
\end{eqnarray}
where $A=87$ is the number of nucleons of each Rubidium atom, $\lambda$ and $r_C$ are the CSL  parameters, and $\hat{\ve{x}}_{\alpha}$ the position operator of the $\alpha$-th atom. In the second line of Eq.~\eqref{lcsl} we performed a Fourier transform which  simplifies the structure of the master equation in Eq.~\eqref{cslmasteq} and will highlight the connection with the non-Markovian and dissipative master equations. By inserting the second line of Eq.~\eqref{lcsl} in  Eq.~\eqref{cslmasteq} and performing the integration over $\ve y$ one obtains:
\begin{equation}
\label{cslmasteq2}
\begin{split}
&\frac{d\hat{\rho}(t)}{dt}=-\frac{i}{\hbar}\left[\hat{H},\hat{\rho}(t)\right]+\frac{\lambda A^{2}r_{C}^{3}}{(\sqrt{\pi}\hbar)^{3}}\sum_{\alpha,\beta=1}^{N}\,\int d\mathbf{Q}\,e^{-\frac{r_{C}^{2}}{\hbar^{2}}\mathbf{Q}^{2}} \\
&\times\left(e^{\frac{i}{\hbar}\mathbf{Q}\cdot\hat{\mathbf{x}}_{\alpha}}\hat{\rho}(t)e^{-\frac{i}{\hbar}\mathbf{Q}\cdot\hat{\mathbf{x}}_{\beta}}-\frac{1}{2}\left\{ \,e^{-\frac{i}{\hbar}\mathbf{Q}\cdot\hat{\mathbf{x}}_{\beta}}\,e^{\frac{i}{\hbar}\mathbf{Q}\cdot\hat{\mathbf{x}}_{\alpha}},\hat{\rho}(t)\right\} \right).
\end{split}
\end{equation}

Given a generic observable $\hat{O}$, the equation for its expectation value $\langle \hat{O} \rangle_t \equiv \textrm{Tr}\{\hat{\rho} (t) \hat{O}\}$ is:
\begin{equation}
\label{cslaverage}
\begin{split}
&\frac{d \langle \hat{O} \rangle_t}{dt} = -\frac{i}{\hbar} \textrm{Tr} \left\{\hat{\rho} (t) \left[\hat{O}, \hat{H}\right] \right\} +\frac{\lambda A^{2}r_{C}^{3}}{(\sqrt{\pi}\hbar)^{3}} \\
&\times\sum_{\alpha,\beta=1}^{N}\,\int d\mathbf{Q}\,e^{-\frac{r_{C}^{2}}{\hbar^{2}}\mathbf{Q}^{2}}\left\{ \textrm{Tr}\left(\hat{\rho}(t)e^{-\frac{i}{\hbar}\mathbf{Q}\cdot\hat{\mathbf{x}}_{\beta}}Oe^{\frac{i}{\hbar}\mathbf{Q}\cdot\hat{\mathbf{x}}_{\alpha}}\right)\right. \\
&\left.-\frac{1}{2}\textrm{Tr}\left(\hat{\rho}(t)\left\{ O,e^{-\frac{i}{\hbar}\mathbf{Q}\cdot\hat{\mathbf{x}}_{\beta}}\,e^{\frac{i}{\hbar}\mathbf{Q}\cdot\hat{\mathbf{x}}_{\alpha}}\right\} \right)\right\}.
\end{split}
\end{equation}
We are interested in the case where $\hat{O}$ can be written as the sum of single-atom observables $\hat{O}_\gamma$,
\begin{equation}
\label{ourop}
\hat{O} = \sum_{\gamma=1}^{N} \, \hat{O}_\gamma.
\end{equation}
In such a case, it is easy to show that when the term $O_{\gamma}$ of the sum in Eq.~\eqref{ourop} is considered, only the Lindblad terms of Eq.~\eqref{cslaverage} with indices $\alpha=\beta=\gamma$ give a non-vanishing contribution. This, together with the fact that the Hamiltonian is separable, allows to reduce the $N$-atom problem to the single-atom case, \textit{i.e.} we can consider the equation
\begin{equation}
\label{cslaverage2}
\begin{split}
&\frac{d \langle \hat{O}_\gamma \rangle_t}{dt} =-\frac{i}{\hbar} \textrm{Tr} \left[\hat{\rho} (t) \left[\hat{O}_\gamma, \hat{H}_\gamma\right] \right] +\frac{\lambda A^{2}r_{C}^{3}}{(\sqrt{\pi}\hbar)^{3}} \\
& \times\int d\mathbf{Q}\,e^{-\frac{r_{C}^{2}}{\hbar^{2}}\mathbf{Q}^{2}}\textrm{Tr}\left[\hat{\rho}(t)\left(e^{-\frac{i}{\hbar}\mathbf{Q}\cdot\hat{\mathbf{x}}_{\gamma}}O_{\gamma}e^{\frac{i}{\hbar}\mathbf{Q}\cdot\hat{\mathbf{x}}_{\gamma}}-O_{\gamma}\right)\right].
\end{split}
\end{equation}
The quantities we need to compute are: the average position variance
\begin{equation}
\label{defposvar}
\langle \hat{\ve{X}}^2 \rangle_t \equiv \frac{1}{N} \sum_{\gamma = 1}^{N} \, \left( \langle \hat{\ve{x}}_{\gamma}^{2} \rangle_t - \langle \hat{\ve{x}}_{\gamma}\rangle_{t}^{2} \right),
\end{equation}
the average momentum variance
\begin{equation}
\label{defmomvar}
\langle \hat{\ve{P}}^2\rangle_t \equiv \frac{1}{N} \sum_{\gamma = 1}^{N} \, \left( \langle \hat{\ve{p}}_{\gamma}^{2} \rangle_t - \langle \hat{\ve{p}}_{\gamma}\rangle_{t}^{2} \right)
\end{equation}
and the average position-momentum correlation
\begin{equation}
\label{defcorr}
\langle \hat{\ve{X}} \hat{\ve{P}} +  \hat{\ve{P}} \hat{\ve{X}} \rangle_t \equiv \frac{1}{N} \sum_{\gamma = 1}^{N} \,  \langle \hat{\ve{x}}_{\gamma}\hat{\ve{p}}_{\gamma}+\hat{\ve{p}}_{\gamma}\hat{\ve{x}}_{\gamma} \rangle_t.
\end{equation}
Since all  atoms are identical and are in the same initial state, the average quantities simply correspond to the expectation values for a single atom, which is what we will focus on, in the following. Taking $\hat{O}_\gamma=\hat{\ve{x}},\, \hat{\ve{p}} $ in Eq.~\eqref{cslaverage2}, it is straightforward to prove that
\begin{equation}
\label{posmomaverage}
\langle \hat{\ve{x}}\rangle_{t} = \langle \hat{\ve{p}}\rangle_{t} = 0
\end{equation}
i.e. CSL  does not affect the average motion in position and momentum of the atoms. However, the same is not true for the standard deviations. In fact, taking $\hat{O}_\gamma=\hat{\ve{x}}^{2}, \hat{\ve{p}}^{2}$  in Eq.~\eqref{cslaverage2}, one finds that:
\begin{eqnarray}
\label{x2manyder}
\frac{d \langle  \hat{\ve{x}}^{2} \rangle_t}{dt} & = &  \frac{1}{m} \langle \hat{\ve{x}} \cdot \hat{\ve{p}}+  \hat{\ve{p}} \cdot \hat{\ve{x}}\rangle_t, \\
\label{p2harm}
\frac{d \langle \hat{\ve{p}}^{2} \rangle_t}{dt} & = &  \frac{ 3 \lambda A^2 \hbar^2}{2 r_{c}^{2}} - m\omega^2 \langle   \hat{\ve{x}}\cdot\hat{\ve{p}} + \hat{\ve{p}}\cdot\hat{\ve{x}}\rangle_t.
\end{eqnarray}
In a similar way, one can show that the position-momentum correlation satisfies the equation:
\begin{equation}
\label{xpharmonic}
\frac{d \langle   \hat{\ve{x}}\cdot\hat{\ve{p}}+\hat{\ve{p}}\cdot\hat{\ve{x}} \rangle_t}{dt} = \frac{2}{m} \langle \hat{\ve{p}}^{2}\rangle_t -2 m \omega^2 \langle \hat{\ve{x}}^{2} \rangle_{t}.
\end{equation}
%
% while the equation for $\hat{\ve{p}}^{2}$ is
%
\\
The set of first order differential equations~\eqref{x2manyder}--\eqref{xpharmonic} can be solved exactly, the solution being:
\begin{eqnarray}
\langle\hat{\ve{x}}^{2}\rangle_{t}&=&\langle\hat{\ve{x}}^{2}\rangle_{t_{0}}+\frac{1}{2\omega m}\Bigl[\mathcal{B}(\omega) \sin\left(2\omega (t-t_0)\right)-  \label{x2sol} \\ 
&& \mathcal{A}(\omega) \left(1- \cos\left(2\omega (t -t_0)\right) \right) \Bigr], \nonumber\\
\nonumber\\
\langle\hat{\ve{p}}^{2}\rangle_{t}&=&\langle\hat{\ve{p}}^{2}\rangle_{t_{0}}+m\omega^{2} \mathcal{C}(\omega)(t-t_{0})-\frac{m\omega}{2} \Bigl[ \mathcal{B}(\omega)  \label{p2sol}\\
%&-& \frac{m\omega}{2} \left[ B(\omega) \sin\left(2\omega (t-t_0)\right)\right. \nonumber\\
%\nonumber\\
&\times& \sin\left(2\omega (t-t_0)\right)-\mathcal{A}(\omega) \left(1- \cos\left(2\omega (t -t_0)\right) \right)\Bigr]\nonumber
\end{eqnarray}
and
\begin{eqnarray}
\langle\hat{\ve{x}}\cdot\hat{\ve{p}}+\hat{\ve{p}}\cdot\hat{\ve{x}}\rangle_{t}&=&\mathcal{A}(\omega)\sin \left(2\omega (t-t_0)\right)+\label{xpsol}\\
&+& \mathcal{B}(\omega)\cos\left(2\omega (t-t_0)\right)+ \mathcal{C}(\omega),\nonumber
\end{eqnarray}
where the real parameters $\mathcal{A}(\omega),\,\mathcal{B}(\omega),\,\mathcal{C}(\omega),$ are fixed by the initial conditions of the system at the initial time $t= t_0$:
\begin{eqnarray}
\mathcal{A}(\omega)&=&m\omega\langle\hat{\ve{x}}^{2}\rangle_{t_{0}}-\frac{\langle\hat{\ve{p}}^{2}\rangle_{t_{0}}}{m\omega}, \nonumber \\
\mathcal{B}(\omega)&=&\langle\hat{\ve{x}}\cdot\hat{\ve{p}}+\hat{\ve{p}}\cdot\hat{\ve{x}}\rangle_{t_{0}}-\mathcal{C}(\omega), \nonumber\\
\nonumber\\
\mathcal{C}(\omega)&=&\frac{3\lambda A^{2}\hbar^{2}}{2 mr_{C}^{2}\omega^{2}}
\end{eqnarray}
The free evolution (i.e. without the harmonic trap) for $\langle\hat{\ve{x}}^{2}\rangle_{t}$, $\langle\hat{\ve{p}}^{2}\rangle_{t}$ and $\langle   \hat{\ve{x}}\cdot\hat{\ve{p}}+\hat{\ve{p}}\cdot\hat{\ve{x}} \rangle_t$ can be obtained by taking the limit $\omega\rightarrow 0$ in Eqs.~\eqref{x2sol}, \eqref{p2sol} and \eqref{xpsol}. In such a case we have:    
\begin{eqnarray}
\langle\hat{\ve{x}}^{2}\rangle_{t}&=&\langle\hat{\ve{x}}^{2}\rangle_{t_{0}}+\frac{\langle\hat{\ve{x}}\cdot\hat{\ve{p}}+\hat{\ve{p}}\cdot\hat{\ve{x}}\rangle_{0}}{m}(t-t_{0})+ \label{x2free} \\
&+&\frac{\langle\hat{\ve{p}}^{2}\rangle_{t_{0}}}{m^{2}}(t-t_{0})^{2}+\frac{\lambda A^{2}\hbar^{2}}{2m^{2}r_{C}^{2}}(t-t_{0})^{3},\nonumber\\
\nonumber\\
\langle\hat{\ve{p}}^{2}\rangle_{t}&=&\langle\hat{\ve{p}}^{2}\rangle_{t_{0}}+\frac{3\lambda A^{2}\hbar^{2}}{2r_{C}^{2}}(t-t_{0}),\label{p2free}
\end{eqnarray}
and for the correlation
\begin{eqnarray}
\langle\hat{\ve{x}}\cdot\hat{\ve{p}}+\hat{\ve{p}}\cdot\hat{\ve{x}}\rangle_{t}&=&\langle\hat{\ve{x}}\cdot\hat{\ve{p}}+\hat{\ve{p}}\cdot\hat{\ve{x}}\rangle_{t_0}+\label{xpfree}\\
&+&\frac{2\langle\hat{\ve{p}}^{2}\rangle_{t_{0}}}{m}(t-t_{0})+\frac{3\lambda A^{2}\hbar^{2}}{2mr_{C}^{2}}(t-t_{0})^{2}.\nonumber
\end{eqnarray}
Given the above equations, we can  easily compute the evolution of $\langle \hat{\ve{x}}^{2} \rangle_{t}$ during the experiment. From $t=0$ to $t=t_1$ the system evolves freely ($\omega=0$) accordingly to Eqs.~\eqref{x2free}--\eqref{xpfree};  from $t=t_1$ to time $t=t_2$ it evolves harmonically as described in Eqs.~\eqref{x2sol}--\eqref{xpsol} and then again freely up to  time $t= t_3$. Imposing the continuity condition during the whole process, one arrives at the final result: 
\begin{equation}
\label{posvart2}
\langle \hat{\ve{x}}^2 \rangle_{ t_3} =  \langle \hat{\ve{x}}^2 \rangle_{ t_3}^{\text{\tiny{QM}}} + \langle \hat{\ve{x}}^2 \rangle_{ t_3}^{\text{\tiny{CSL}}} ,
\end{equation}
where $\langle \hat{\ve{x}}^2 \rangle_{t}^{\text{\tiny{QM}}}$ is the value of the position variance according to the standard Schr\"odinger evolution, and $\langle \hat{\ve{x}}^2 \rangle_{t}^{\text{\tiny{CSL}}}$ is the modification induced by CSL. The first term has the  form
\begin{eqnarray}\label{posvart2qm}
\langle \hat{\ve{x}}^2 \rangle_{ t_3}^{\text{\tiny{QM}}} &=& A^{\text{\tiny{QM}}} (\omega,t_1,t_3,\delta t_2)%+B^{\text{\tiny{QM}}}  (\omega,t_1,t_3,\delta t_2) \cos (2\omega \delta t_2) 
\\
\nonumber\\
&+& B^{\text{\tiny{QM}}}  (\omega,t_1,t_3,\delta t_2) \cos (2\omega \delta t_2)\nonumber\\
\nonumber\\
&+&  C^{\text{\tiny{QM}}}  (\omega,t_1,t_3,\delta t_2) \sin (2\omega \delta t_2),\nonumber
\end{eqnarray}
where we defined the following quantity:
\begin{widetext}
\begin{subequations}
\label{parqm}
\begin{align}
&A^{\text{\tiny{QM}}} = \frac{\left[\langle \hat{\ve{p}}^2 \rangle_{0} + \left( \langle \hat{\ve{x}}^2 \rangle_{0} m^2 + \langle \hat{\ve{p}}^2 \rangle_{0} t_{1}^{2}\right)\omega^2\right] \left[1 +\left(  t_3 - t_2 \right)^2 \omega^2\right] }{2 m^2 \omega^2}; \\
& B^{\text{\tiny{QM}}}  = - \frac{\langle \hat{\ve{p}}^2 \rangle_{0} - \left[\langle \hat{\ve{x}}^2 \rangle_{0} m^2 +\langle \hat{\ve{p}}^2 \rangle_{0}\left(   \left(  t_3 -t_2\right)^2 + 4 t_1 \left(  t_3 - t_2\right) + t_{1}^{2} \right)\right]\omega^2 }{2 m^2 \omega^2} +  \notag \\
&\frac{ \left(\langle \hat{\ve{x}}^2 \rangle_{0} m^2 + \langle \hat{\ve{p}}^2 \rangle_{0}t_{1}^{2} \right) \left(  t_3 - t_2 \right)^2 \omega^2}{2 m^2};\\
& C^{\text{\tiny{QM}}} = \frac{\langle \hat{\ve{p}}^2 \rangle_{0} \left(  t_2-\tau_p\right) - \left[\langle \hat{\ve{x}}^2 \rangle_{0} m^2 + \langle \hat{\ve{p}}^2 \rangle_{0} t_1 ( t_2 - \tau_p) \right] \left(  t_3 - t_2\right)\omega^2 }{\omega m^2}.
\end{align}
\end{subequations}
The CSL contribution is given by 
\begin{equation}\label{posvart2csl}
\langle \hat{\ve{x}}^2 \rangle_{ t_3}^{\text{\tiny{CSL}}} = \frac{\lambda A^2 \hbar^2}{r_{C}^{2} 8 m^2 \omega^3} \Bigl [A^{\text{\tiny{CSL}}} (\omega,t_1,t_3,\delta t_2) + B^{\text{\tiny{CSL}}}  (\omega,t_1,t_3,\delta t_2) \cos (2\omega \delta t_2) +  C^{\text{\tiny{CSL}}}  (\omega,t_1,t_3,\delta t_2) \sin (2\omega \delta t_2)\Bigr]
\end{equation}
with
\begin{subequations}
\label{parcsl}
\begin{align}
&A^{\text{\tiny{CSL}}} =6 \omega  t_3 + 2 \omega^3 \left[t_{2}^3 + 2 t_{3}^{3} + t_{1}^{3}  - 3 t_{3}^{2} t_{2}\right]+ 2 t_{1}^{3} \left(  t_3 - t_2\right)^2 \omega^5; \\
& B^{\text{\tiny{CSL}}}  = -2 \omega \left[3 ( t_3 -\delta t_2) + \omega^2 t_1 (2 t_{1}^{2} - 3 ( t_3 -\delta t_2)^2) +\omega^4 t_{1}^{3} ( t_3 -t_2)^2 \right]; \\
& C^{\text{\tiny{CSL}}} = 3 +3 \omega^2 \left[ ( t_3-t_2)^2-2 ( t_3-\delta t_2)^2\right]+ 2\omega^4 t_{1}^{2}( t_3-t_2)(3  t_3 -t_1 - 3\delta t_2).
\end{align}
\end{subequations} 
\end{widetext}
We can see that the CSL contribution $\langle \hat{\ve{x}}^2 \rangle_{ t_3}^{\text{\tiny{CSL}}}$ to the final variance is independent from the initial state of the gas (contrary to $\langle \hat{\ve{x}}^2 \rangle_{ t_3}^{\text{\tiny{QM}}}$) and depends, in a rather complicated way, only on the relevant times of the experiment ($t_1$, $\delta t_2=t_2-t_1$, $t_3$) and the frequency $\omega$ of the delta-kick. 
\\

\begin{figure}[t]
\includegraphics[scale=0.7]{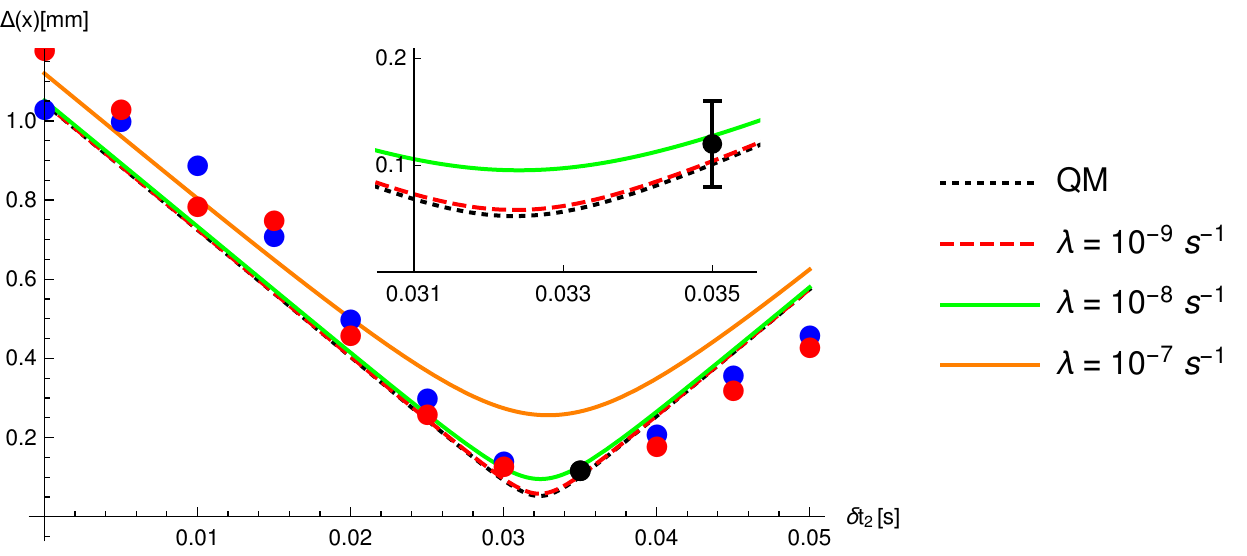}
\caption{Position's standard deviation $\Delta (x) \equiv \langle \hat{\ve{x}}^2 \rangle_{t_3}^{1/2}$ at the detector (time $t = t_3$) as a function of the delta-kick time $\delta t_2$, for three different values of the collapse rate $\lambda$. For each curve, we fixed $r_C = 10^{-7}$ m. The inset shows the curves near the minimum value detected in~\cite{kas}, $\Delta (x)_{\text{\tiny EXP}} = 120^{+40}_{-40} \mu$m, indicated by the black bars. The black dotted line shows the quantum-mechanical predictions. The red and blue points represent the experimental data deducted from Fig.~$3$ in~\cite{kas}.}
\label{pos-taup}
\end{figure}

In Fig.~\ref{pos-taup} we plot  the final position variance $\langle \hat{\ve{x}}^2 \rangle_{t_3}$ as a function of the delta-kick time $\delta t_2$. To highlight the CSL effect, we computed the quantum-mechanical prediction and the CSL predictions for three different values of $\lambda$ and at fixed $r_C=10^{-7}$ m.

As we can see, for small  $\lambda$ the quantum-mechanical predictions, compatible with the experimental data, are recovered. For larger values of $\lambda$ the  variance $\langle \hat{\ve{x}}^2 \rangle_{t_3}$ increases, till it disagrees with the experimental data. This is the expected behavior: the larger $\lambda$, the stronger the Brownian fluctuations and the larger the spread of the cloud. 

\begin{figure}[t]
\includegraphics[scale=0.7]{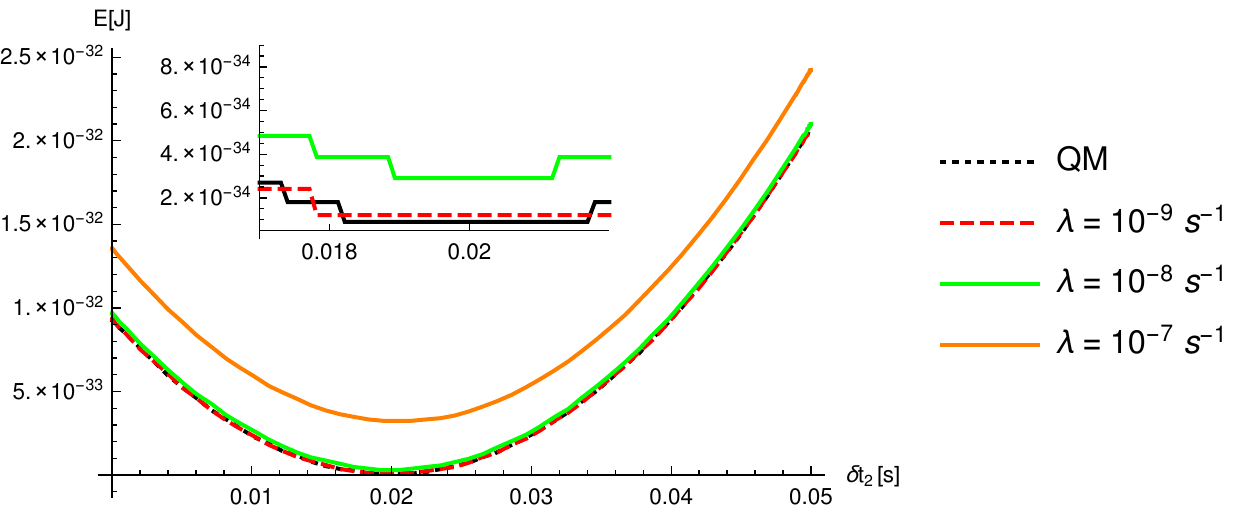}
\caption{Kinetic energy $E \equiv \langle \hat{\ve{p}}^2 \rangle_{t_3}/ 2m $ at the detector (time $t = t_3$) as a function of the delta-kick time $\delta t_2$, for three different values of the collapse rate $\lambda$. For each curve, we fixed $r_C = 10^{-7}$ m. The inset shows the minimum of the curves, which is $E \sim 10^{-34}$J, corresponding to a temperature $T \simeq 10$ pK, for $\delta t_2 \approx 20$ ms. The black dotted line shows the quantum-mechanical predictions.}
\label{kin-taup}
\end{figure}

Similarly, in Fig.~\ref{kin-taup} we plot the average energy of the gas at the end of the process as a function of $\delta t_2 $, for different values of $\lambda$ and again at fixed $r_C=10^{-7}$ m. We see that the cooling effect is maximum when the delta-kick last for $\delta t_2\approx 20$ ms, leading to a theoretical kinetic energy of $E \approx 10^{-34}$ J, corresponding to a temperature of order $T \approx 10$ pK. This theoretical value is compatible with the experimental value $T_{min} = 50_{-30}^{+50}$ pK measured in ~\cite{kas}. We also note that the heating effect due to CSL becomes significant for $\lambda \geq 10^{-7} \, \text{s}^{-1}$, leading to an energy increase greater than $ 5 \times 10^{-33}$ J, which is about $5$ times greater than the value of the energy increase measured during the experiment ($(4\pm6)\times 10^{-34}$ J).
%
%%%%%%%%%%%%%%%%%%%%%%%%%%%%%%%%%%%%%%%%%%%%%%%%%%%%%%%%%%%%%%%%%%%%%
\section{Expansion of the gas according to the non-white CSL model}\label{color}
%%%%%%%%%%%%%%%%%%%%%%%%%%%%%%%%%%%%%%%%%%%%%%%%%%%%%%%%%%%%%%%%%%%%%%
We now consider the predictions of CSL with a non-white noise (cCSL) on the expansion of the gas. 
%
%A non-white noise model is, in fact, more realistic than a white-noise model (corresponding to an infinite energy source). Therefore it's important to check the robustness of the previous analysis also in presence of a more general and realistic non-white noise. 
%
The (single particle) cCSL master equation~\cite{nonwhite1, nonwhite2}, to the first perturbative order in $\lambda$, is~\footnote{Here we report only the single-atom master equation because, similarly to the case of white noise CSL model, we only need to focus on single-atom observables in order to describe the gas.}:
\begin{eqnarray}\label{ro_non-white}
\frac{d\rho\left(t\right)}{dt} &=& -\frac{i}{\hbar}\left[H,\rho\left(t\right)\right]-\lambda8\pi^{3/2}r_{C}^{3}A^{2}\int_{0}^{t}ds \,f\left(s\right)\\
\nonumber\\
&\times&\int d\mathbf{Q}\tilde{g}(\mathbf{Q})\tilde{g}(-\mathbf{Q})\left[e^{-\frac{i}{\hbar}\mathbf{Q}\cdot\hat{\mathbf{x}}},\left[e^{\frac{i}{\hbar}\mathbf{Q}\cdot\hat{\mathbf{x}}(-s)},\rho\left(t\right)\right]\right]\nonumber
\end{eqnarray}
where
\begin{equation}\label{g_k}
\tilde{g}(\mathbf{Q})=\frac{1}{\left(2\pi\hbar\right)^{3/2}}e^{-\frac{\mathbf{Q}^{2}r_{C}^{2}}{2\hbar^{2}}},
\end{equation}
the function $f\left(s\right)$ is the time correlation function of the non-white noise,  and $\hat{\mathbf{x}}(-s)$ is the position operator in the interaction picture, evolved backwards to the time $-s$:  
\begin{equation}
\hat{\mathbf{x}}(-s)=e^{-\frac{i}{\hbar}H s}\,\hat{\mathbf{x}}\,e^{\frac{i}{\hbar}H s}.
\end{equation}
In the white noise limit the correlation function  $f(s)$ becomes a Dirac-delta and the standard CSL master equation~\eqref{cslmasteq2} with $N=1$ is recovered.

From Eq.~\eqref{ro_non-white} it is easy to derive the evolution equation for a generic operator $O$:
\begin{eqnarray}\label{non-white}
\frac{d\langle \hat{O}\rangle _{t}}{dt} &=& -\frac{i}{\hbar}\langle [\hat{O},\hat{H}]\rangle _{t}-\lambda8\pi^{3/2}r_{C}^{3} A^{2}\int_{0}^{t}ds f\left(s\right)\\
\nonumber\\
&\times&\int d\mathbf{Q}\tilde{g}(\mathbf{Q})\tilde{g}(-\mathbf{Q})\langle [[\hat{O},\,e^{-\frac{i}{\hbar}\mathbf{Q}\cdot\hat{\mathbf{x}}}],\,e^{\frac{i}{\hbar}\mathbf{Q}\cdot\hat{\mathbf{x}}(-s)}]\rangle _{t}. \nonumber
\end{eqnarray}
This non-Markovian master equation cannot be solved exactly for a general non-white noise. We can proceed as follows, noting that any realistic correlation function has a cut-off time $\tau$ (to which a frequency cut-off $\Omega$ corresponds). When $\tau$ is much smaller than the typical  timescales of the system, the new dynamics is expected to be indistinguishable from the white-noise case. We will assess for which values of $\tau$ the white noise limit is recovered. More precisely we are interested in determining when we can approximate: 
\begin{equation}\label{approx}
e^{\frac{i}{\hbar}\mathbf{Q}\cdot\hat{\mathbf{x}}(-s)}\simeq e^{\frac{i}{\hbar}\mathbf{Q}\cdot\hat{\mathbf{x}}}. 
\end{equation}

Given the harmonic  Hamiltonian in  Eq.~\eqref{harmonic}, the position operator in the interaction picture evolves as follows:
\begin{equation}\label{x_int_pic}
\hat{\ve{x}}\left(-s\right)=\cos\left(\omega s\right)\hat{\ve{x}}-\frac{\sin\left(\omega s\right)}{m\omega}\hat{\ve{p}}
\end{equation}
which implies 
\begin{equation} \label{eq:hrty}
e^{\frac{i}{\hbar}\mathbf{Q}\cdot\hat{\mathbf{x}}(-s)}=e^{\frac{i}{\hbar}\cos(\omega s)\mathbf{Q}\cdot\hat{\ve{x}}}\,e^{-\frac{i}{\hbar}\frac{\sin(\omega s)}{m\omega}\mathbf{Q}\cdot\hat{\ve{p}}}\,e^{-\frac{i}{\hbar}\frac{\sin(2\omega s)}{4m\omega}\mathbf{Q}^{2}}.
\end{equation}
We perform the analysis under the assumption that 
\begin{equation}
\label{nwcond0}
\tau \ll t\simeq 10^{-2}\;\textrm{s},
\end{equation}
which is the order of magnitude of the delta-kick time. This assumption is necessary in order to obtain conditions, which depend only on the noise cut-off $\tau$ and not on the time $t$ of evolution. Then, according to Eq.~\eqref{eq:hrty}, the approximation in Eq.~\eqref{approx} is fulfilled when:
\begin{equation}
\label{nwcond1}
\omega \tau\ll1 \;\;\;\Rightarrow\;\;\; \tau\ll \omega^{-1}\simeq 0,94 \textrm{s},
\end{equation}
and:
\begin{equation}
\label{nwcond2}
\frac{|\ve{Q}||\ve{p}_{max}|\tau}{\hbar m} \ll 1\;\;\;\Rightarrow\;\;\; \tau\ll 10^{3} \left( \frac{r_C}{1 \textrm{m}}\right) \textrm{s},
\end{equation}
and also:
\begin{equation}
\label{nwcond3}
\frac{ \tau}{2m \hbar}\mathbf{Q}^{2}\ll1\;\;\;\Rightarrow\;\;\;\tau\ll 10^{9} \left( \frac{r_{C}^{2}}{1 \textrm{m}^2}\right) \textrm{s},
\end{equation}
where $m=1,44\times10^{-25}$ Kg is the Rb mass, the maximum momentum is $|\mathbf{p}_{max}|=\left\langle \hat{\mathbf{p}}\right\rangle +\left\langle \hat{\mathbf{p}}^{2}\right\rangle^{1/2}\simeq 10^{-29}$ Kg m/s (we took $\langle \hat{\mathbf{p}}\rangle =0$ and $\langle E\rangle =\left\langle \hat{\mathbf{p}}^{2}/2m\right\rangle \simeq10^{-32}$ J) and $|\mathbf{Q}|\leq \hbar/r_{C}$, which is imposed by the Gaussian factors $\tilde{g}(\mathbf{Q})$ defined in Eq.~\eqref{g_k}.
%For small values of $r_C$, conditions in Eq.~\eqref{nwcond2} and Eq.~\eqref{nwcond3} are in general quite strong. For example, %if $r_C = 10^{-7}$ m, they are satisfied with $\tau \leq 10^{-5}$ s.

%Taking into account that, in our case $t \approx 1$ s for the free evolution, and $t\approx 35$ ms for the harmonic evolution, from the conditions in Eqs.~\eqref{nwcond0} to~\eqref{nwcond3} this approximation holds for $\tau \ll 10^{-2}$ s for $r_C \geq 10^{-5}$ m, coming from the strongest bound in Eq.~\eqref{nwcond0}; for lower values of $r_C$ the strongest bound comes from conditions in Eqs.~\eqref{nwcond2}  and~\eqref{nwcond3}.

Given the assumption in Eq.~\eqref{nwcond0}, the condition in Eq.~\eqref{nwcond1} is always fulfilled, as well as conditions in Eqs.~\eqref{nwcond2} and \eqref{nwcond3}, as long as $r_C \geq 10^{-5}$ m. On the other hand, for $r_C \leq 10^{-5}$ m, the strongest bound comes from the conditions in Eqs.~\eqref{nwcond2} and~\eqref{nwcond3}.

Under these conditions, the evolution equation for a generic operator $O$ becomes:
\begin{eqnarray}
\frac{d\langle \hat{O}\rangle _{t}}{dt} &=& -\frac{i}{\hbar}\langle [\hat{O},\hat{H}]\rangle _{t}-\frac{\lambda8\pi^{3/2}r_{C}^{3} A^{2}\tilde{f}(0)}{2}\\
\nonumber\\
&\times&\int d\mathbf{Q}\tilde{g}(\mathbf{Q})\tilde{g}(-\mathbf{Q})\langle [[\hat{O},\,e^{-\frac{i}{\hbar}\mathbf{Q}\cdot\hat{\mathbf{x}}}],\,e^{\frac{i}{\hbar}\mathbf{Q}\cdot\hat{\mathbf{x}}}]\rangle _{t}\nonumber
\end{eqnarray}
where
\begin{equation}
\tilde{f}(\omega):=\int_{-\infty}^{+\infty}f(s)e^{i\omega s}ds
\end{equation}
and where we assumed $f(s)=f(-s)$ and used the fact that for $t>\tau$ 
\begin{equation}
\int_{0}^{t}ds f\left(s\right)\simeq \int_{0}^{\infty}ds f\left(s\right)=\frac{1}{2}\int_{-\infty}^{\infty}ds f\left(s\right)=\frac{\tilde{f}(0)}{2}.
\end{equation}
Therefore, under the assumption Eq.~\eqref{nwcond0} and when conditions Eq.~\eqref{nwcond2} and Eq.~\eqref{nwcond3} are fulfilled, the non-white noise case is well approximated by the  white-noise case discussed in the previous section, with the replacement $\lambda \rightarrow \lambda \tilde{f}(0)/2$. 

A more detailed analysis is possible for a system with spatial extension smaller than $r_C$. In our case $\left \langle \hat{\ve{x}}^2 \right \rangle^{1/2} \approx 50 \, \mu\textrm{m}$, implying that the approximation holds for $r_C \geq 10^{-4}$ m. Imposing this condition on the Gaussian factors $\tilde{g}(\mathbf{Q})$ defined in Eq.~\eqref{g_k} gives $|\ve{Q}|\leq \hbar/r_{C}$, which guarantees that we can expand the exponentials in the second line of Eq.~\eqref{non-white} as $e^{-\frac{i}{\hbar}\mathbf{Q}\cdot\hat{\mathbf{x}}}\simeq 1-\frac{i}{\hbar}\mathbf{Q}\cdot\hat{\mathbf{x}}$, leading to
\begin{equation}
\sum_{i,j=1}^{3}\left(\frac{1}{\hbar^2}\int d\mathbf{Q}\tilde{g}(\mathbf{Q})\tilde{g}(-\mathbf{Q})Q_{i}Q_{j}\right)\langle [[\hat{O},\,\hat{x}_{i}],\,\hat{x}_{j}(-s)]\rangle _{t}.
\end{equation}
The integration over $\ve{Q}$ gives the factor
\begin{equation}
\frac{1}{\hbar^2}\int d\mathbf{Q}\tilde{g}(\mathbf{Q})\tilde{g}(-\mathbf{Q})Q_{i}Q_{j}=\frac{\delta_{ij}}{2^{4}\pi^{3/2}r_{C}^{5}} 
\end{equation}  
and therefore Eq.~\eqref{non-white} becomes
\begin{eqnarray}\label{masteqnonwhite}
\frac{d\langle \hat{O}\rangle _{t}}{dt} &=& -\frac{i}{\hbar}\langle [\hat{O},\hat{H}]\rangle _{t}-\frac{\lambda A^{2}}{2 r_{C}^{2}}\int_{0}^{t}ds f\left(s\right)\;\;\\
\nonumber\\
&\times&\sum_{j=1}^{3}\langle [[\hat{O},\,\hat{x}_{j}],\,\hat{x}_{j}(-s)]\rangle _{t}. \nonumber
\end{eqnarray}

An explicit calculation is also possible, if we take a specific expression for  the noise correlator, e.g.:
\begin{equation}
\label{f}
f (s) = \frac{1}{2\tau} e^{-\abs{s}/\tau}.
\end{equation}
which, in the limit $\tau \rightarrow 0$, reduces to a Dirac delta.
From Eq.~\eqref{masteqnonwhite} it is easy to see that the dynamical equations for $\hat{\ve{x}}$ and $\hat{\ve{p}}$ are not modified by the noise. Similarly, for $\hat{\ve{x}}^2$ we have:
\begin{equation}
\label{x2nonwhite}
\frac{d\left\langle \hat{\ve{x}}^2\right\rangle _{t}}{dt} = \frac{\left\langle\hat{\ve{x}} \hat{\ve{p}} + \hat{\ve{p}}\hat{\ve{x}} \right\rangle_{t}}{m}.
\end{equation}
From Eq.~\eqref{masteqnonwhite}, it is also straightforward obtain the following equations:
\begin{eqnarray}\label{xppxnonwhite}
\frac{d\left\langle\hat{\ve{x}} \hat{\ve{p}} + \hat{\ve{p}}\hat{\ve{x}}\right\rangle _{t}}{dt} &=& \frac{2\left\langle \hat{\ve{p}}^2\right\rangle _{t}}{m} - 2 m \omega^2 \left\langle \hat{\ve{x}}^2\right\rangle _{t} \nonumber\\
\nonumber\\
&+& \frac{3 \lambda A^2 \hbar^2}{m r_{C}^{2}}\int_{0}^{t} ds \, \frac{e^{-\frac{s}{\tau}} \sin (s \omega)}{2 \omega \tau};\\\nonumber
\end{eqnarray} 
\begin{equation}
\label{p2nonwhite}
\frac{d\left\langle \hat{\ve{p}}^2\right\rangle _{t}}{dt}= -m\omega^2 \left\langle\hat{\ve{x}} \hat{\ve{p}} + \hat{\ve{p}}\hat{\ve{x}} \right \rangle_{t} + \frac{3 \lambda A^2 \hbar^2}{2 r_{C}^{2}}\int_{0}^{t} ds \, \frac{e^{-\frac{s}{\tau}} \cos (s \omega)}{2 \tau}.
\end{equation}
The system of Eqs.~\eqref{x2nonwhite}, \eqref{xppxnonwhite} and \eqref{p2nonwhite} can be solved exactly. The solution of Eq.~\eqref{xppxnonwhite} is:
\begin{widetext}
%\begin{eqnarray}
%\label{xppxnonwhite2}
%\left\langle\hat{\ve{x}} \hat{\ve{p}} + \hat{\ve{p}}\hat{\ve{x}}\right\rangle _{t} &=& \left\langle\hat{\ve{x}} %\hat{\ve{p}} + \hat{\ve{p}}\hat{\ve{x}}\right\rangle _{0} \cos (2 \omega t) \nonumber\\
%\nonumber\\
%&+& \left ( \frac{\left\langle \hat{\ve{p}}^2\right\rangle _{0}}{m \omega} - m \omega \left\langle \hat{\ve{x}}^2\right%\rangle _{0}\right) \sin(2 \omega t)\nonumber\\
%\nonumber\\
%&+& \frac{3 \lambda A^2 \hbar^2}{2\omega m r_{C}^{2}} \int_{0}^{t} ds g(s) \sin (2\omega (t-s)),
%\end{eqnarray}
\begin{equation}
\label{xppxnonwhite2}
\left\langle\hat{\ve{x}} \hat{\ve{p}} + \hat{\ve{p}}\hat{\ve{x}}\right\rangle _{t} = \left\langle\hat{\ve{x}} \hat{\ve{p}} + \hat{\ve{p}}\hat{\ve{x}}\right\rangle _{0} \cos (2 \omega t)+ \left ( \frac{\left\langle \hat{\ve{p}}^2\right\rangle _{0}}{m \omega} - m \omega \left\langle \hat{\ve{x}}^2\right\rangle _{0}\right) \sin(2 \omega t)+ \frac{3 \lambda A^2 \hbar^2}{2\omega m r_{C}^{2}} \int_{0}^{t} ds g(s) \sin (2\omega (t-s)),
\end{equation}
where
\begin{equation}
\label{g}
g (x) =  \int_{0}^{x} dy \, \frac{e^{-\frac{y}{\tau}} \cos(\omega y)}{2\tau} + \frac{e^{-\frac{x}{\tau}} \sin(\omega x)}{2\omega\tau}. 
\end{equation}
Using Eq.~\eqref{xppxnonwhite2} in Eqs.~\eqref{x2nonwhite} and \eqref{p2nonwhite} we get the related solutions:
\begin{equation}
\label{x2nonwhite2}
\begin{split}
&\left\langle \hat{\ve{x}}^2\right\rangle _{t} = \left\langle \hat{\ve{x}}^2\right\rangle _{0} + \frac{1}{2 m \omega}\left [\sin(2 \omega t)  \left\langle\hat{\ve{x}} \hat{\ve{p}} + \hat{\ve{p}}\hat{\ve{x}}\right\rangle _{0} - \left ( \frac{\left\langle \hat{\ve{p}}^2\right\rangle _{0}}{m \omega} -  m \omega \left\langle \hat{\ve{x}}^2\right\rangle _{0}\right) (1 -\cos (2\omega t)) \right] \\
& + \frac{3 \lambda A^2 \hbar^2}{2\omega m^2 r_{C}^{2}} \int_{0}^{t} ds_2\int_{0}^{s_2} ds_1 \, g(s_1) \sin (2\omega (s_2-s_1));
\end{split}
\end{equation}
\begin{equation}
\label{p2nonwhite2}
\begin{split}
&\left\langle \hat{\ve{p}}^2\right\rangle _{t} = \left\langle \hat{\ve{p}}^2\right\rangle _{0} - \frac{m \omega}{2}\left [\sin(2 \omega t)  \left\langle\hat{\ve{x}} \hat{\ve{p}} + \hat{\ve{p}}\hat{\ve{x}}\right\rangle _{0} - \left ( \frac{\left\langle \hat{\ve{p}}^2\right\rangle _{0}}{m \omega} -  m \omega \left\langle \hat{\ve{x}}^2\right\rangle _{0}\right) (1 -\cos (2\omega t)) \right] \\
& +  \ \frac{3 \lambda A^2 \hbar^2}{2 r_{C}^{2}}\left[  \int_{0}^{t} ds_2\int_{0}^{s_2} ds_1 \left( \frac{e^{-\frac{s_1}{\tau}} \cos(\omega s_1)}{2\tau} -\omega g(s_1) \sin (2\omega (s_2-s_1)) \right) \right ].
\end{split}
\end{equation}
\end{widetext}
From a direct computation of the function ~\eqref{g}, it is possible to note that, if $\tau \omega \ll 1$ and $\tau \ll t$, then the solutions Eqs.~\eqref{xppxnonwhite2}, \eqref{x2nonwhite2} and \eqref{p2nonwhite2} are practically indistinguishable from Eqs.~\eqref{x2sol}, \eqref{p2sol} and \eqref{xpsol} derived in the white noise case. In the experiment under consideration, we have $\omega = 6.7$ rad/s and $t = \delta t_2\approx 35$ ms. The white noise limit is therefore a good approximation for any noise with cut-off $\tau \leq 10^{-3}$s.

In the free evolution limit $\omega \to 0$, Eqs.~\eqref{x2nonwhite2}, \eqref{xppxnonwhite2} and \eqref{p2nonwhite2} reduce to:
\begin{equation}
\label{p2nonwhite2free}
\left\langle \hat{\ve{p}}^2\right\rangle _{t} = \left\langle \hat{\ve{p}}^2\right\rangle _{0} + \frac{3\lambda A^2 \hbar^2}{2 r_{C}^2} \left [  t - \tau \left(1 - e^{-\frac{t}{\tau}} \right) \right];
\end{equation}
\begin{eqnarray}
\label{xppxnonwhite2free}
\left\langle\hat{\ve{x}} \hat{\ve{p}} + \hat{\ve{p}}\hat{\ve{x}}\right\rangle _{t} &=& \left\langle\hat{\ve{x}} \hat{\ve{p}} + \hat{\ve{p}}\hat{\ve{x}} \right\rangle_{0} + \frac{2\left\langle \hat{\ve{p}}^2\right\rangle _{0}t}{m} \\
\nonumber\\
&+&\frac{3\lambda A^2 \hbar^2 }{2 m r_{C}^2} \left [ t^2- \tau t \left(1 - e^{-\frac{t}{\tau}}\right)\right];\nonumber
\end{eqnarray}
\begin{eqnarray}
\label{p2nonwhite2free}
&&\left\langle \hat{\ve{x}}^2\right\rangle _{t} = \left\langle \hat{\ve{x}}^2\right\rangle _{0} +  \frac{\left\langle\hat{\ve{x}} \hat{\ve{p}} + \hat{\ve{p}}\hat{\ve{x}} \right\rangle_{0} t}{m} + \frac{\left\langle \hat{\ve{p}}^2\right\rangle _{0} t^2}{ m^2}\\ 
\nonumber\\
&&+\frac{3\lambda A^2 \hbar^2}{m^2 r_{C}^2} \left [\frac{t^3}{6} - \frac{t \tau}{2}\left(\frac{1}{2}+ e^{-\frac{t}{\tau}}\right) +\frac{\tau^3}{2}\left(1 - e^{-\frac{t}{\tau}}\right)\right].\nonumber
\end{eqnarray}
In this case the white noise limit is recovered when $\tau \ll t$. The free time evolution is  $t \approx 1$s, which implies $\tau \leq 10^{-2}$s. 

To conclude, we can safely say that the bounds we obtain for the CSL model shown in Fig.~\ref{exc_csl} hold also for a more general and realistic non-white noise extension of the model if 
\begin{equation}
\tau\leq10^{-3}\,\textrm{s}\quad \Longrightarrow \quad \Omega\geq10^{3}\,\textrm{Hz}
\end{equation}
for $r_{C}\geq10^{-5}\,\textrm{m}$,
\begin{equation}
\tau\ll 10^{3} \left( \frac{r_C}{1 \textrm{m}}\right)\,\textrm{s} \quad \Longrightarrow \quad \Omega\gg10^{-3}\left(\frac{1\textrm{m}}{r_{C}}\right)\,\textrm{Hz}\;\;
\end{equation}
for $10^{-6}\leq r_{C}\leq10^{-5}\,\textrm{m}$,
\begin{equation}
\tau\ll10^{9}\left(\frac{r_{C}^{2}}{1\textrm{m}^{2}}\right)\,\textrm{s} \quad \Longrightarrow \quad \Omega\gg10^{-9}\left(\frac{1\textrm{m}^{2}}{r_{C}^{2}}\right)\,\textrm{Hz}\;\;
\end{equation}
for $r_{C}\leq10^{-6}\,\textrm{m}$. Taking into account that typical cosmological cut-offs are of order $10^{10}-10^{11}$ Hz, our analysis shows that for $r_C \geq 10^{-10}$ m and for a typical cosmological collapse noise, the cCSL predictions (therefore also the upper bounds) are indistinguishable from the standard CSL predictions.  

% Here $\tau$ is the cut-off in time of the correlation function of the noise while $\Omega$ is the corresponding cut-off in the frequency domain. 
  
%%%%%%%%%%%%%%%%%%%%%%%%%%%%%%%%%%%%%%%%%%%%%%%%%%%%%%%%%%%%%%%%%%%%%
\section{Expansion of the gas according to the dCSL  model}
%%%%%%%%%%%%%%%%%%%%%%%%%%%%%%%%%%%%%%%%%%%%%%%%%%%%%%%%%%%%%%%%%%%%%%
%
Another possible way of generalizing the CSL model is offered by the dCSL model~\cite{andrea}, which includes dissipative effects in the dynamics, to tame the energy increase.  More precisely, a finite temperature is associated to the collapse-noise, and every physical system slowly thermalizes to that temperature. 
% As a consequence, the system's initial kinetic energy not only does not increase indefinitely, but in certain cases can even decrease. 
If the noise has a cosmological origin, a temperature of $\sim 1$K is expected, meaning that in general the energy of material objects should actually decrease, not increase as predicted by CSL. Since the effect we are discussing in this paper is directly related to the energy increase, a dissipative modification of CSL is expected to change the bounds on the collapse parameters. This is what we will consider now.

In the dCSL model,  the Lindblad operators $\mathbb{L} (\ve{y})$ are defined as follows:
\begin{equation}
\label{ldcsl}
\begin{split}
\mathbb{L} (\ve{y}) = &\frac{\sqrt{\lambda A^{2}8\pi^{3/2}r_{C}^{3}}}{(2\pi\hbar)^{3}} \sum_{\alpha = 1}^{N} \int d\ve{Q} \, e^{\frac{i}{\hbar}\ve{Q} (\hat{\ve{x}}_{\alpha} - \ve{y})} \times \\
&e^{-\frac{r_{C}^{2}}{\hbar^2}\left \lvert (1+k)\ve{Q} + 2k \hat{\ve{P}}_{\alpha}\right\rvert},
\end{split}
\end{equation}
where the new parameter
\begin{equation}\label{k}
k=\frac{\hbar^2}{8 m k_B T_{\text{\tiny{CSL}}} r_{C}^{2}},
\end{equation}
controls the  temperature $T_{\text{\tiny{CSL}}}$ of the collapse noise. In the limit $T_{\text{\tiny{CSL}}} \rightarrow \infty$ ($k \rightarrow 0$), the standard CSL Lindblad operators of Eq.~\eqref{lcsl} are recovered. 

 As in the case of the standard CSL model, it is easy to prove that for a  gas of non-interacting atoms, the problem can be reduced to the study of single-atom observables. 
%Even if we proved these properties for the CSL model, we will now promote them also to the dissipative case. In fact, as it has been seen in~\cite{andrea}, the dissipation becomes relevant only on the secular evolution of the system, while the localization mechanism is relevant for short time. 
The dCSL model master equation for a single atom trapped in an harmonic potential is given by Eq.~\eqref{cslmasteq}, with  $\mathbb{L}$ defined as in Eq.~\eqref{ldcsl} and with $N=1$. After performing an integration over the variable $\ve y$ we arrive at~\cite{andrea}:
\begin{equation}
\label{dCSL}
\begin{split}
&\frac{d\hat{\rho}}{dt} = -\frac{i}{\hbar} \left[  \frac{\hat{\ve{p}}^2}{2m} +  \frac{1}{2} m \omega^2 \hat{\ve{x}}^{2}, \hat{\rho} \right] + \frac{\lambda A^2  r_{C}^{3}}{(\sqrt{\pi}\hbar)^3}\\
& \int d^3 Q \, \Bigl( e^{\frac{i}{\hbar} \ve{Q}\cdot \hat{\ve{x}}} L(\ve{Q}, \hat{\ve{p}}) \hat{\rho} (t) L(\ve{Q}, \hat{\ve{p}}) e^{-\frac{i}{\hbar} \ve{Q} \cdot \hat{\ve{x}}} \\
& -\frac{1}{2} \left \{  L^2(\ve{Q}, \hat{\ve{p}}),  \hat{\rho} \right \} \Bigr),
\end{split}
\end{equation}
where
\begin{equation}
\label{LdCSL}
L(\ve{Q}, \hat{\ve{p}}) = e^{-\frac{r_{C}^{2}}{2\hbar^2} \left \lvert(1+k)\ve{Q} + 2k\hat{\ve{p}} \right\rvert^2}.
\end{equation}
%
% In the limit $k \to 0$ the standard CSL model is recovered.
% In~\cite{andrea} is proved that
%\begin{equation}
%\label{energydCSL}
%\langle\frac{\hat{p}^2}{2m} \rangle_t = H (t) = e^{-\chi t} \left( H(0) - H_{as} \right) + H_{as},
%\end{equation}
%
%with relaxation rate
%\begin{equation}
%\label{chidCSL}
%\chi = \frac{4k\lambda A^2m^2}{(1+k)^5m_0^2}
%\end{equation}
%
%and the asymptotic energy given by
%\begin{equation}
%\label{has}
%H_{as} = \frac{3\hbar^2}{16km r_{C}^{2}}.
%\end{equation}
%
%The usual expression of CSL model is obtained in the limit $k \to 0$. 

With the help of the above equation, we can easily derive the equation for the variance in position: 
\begin{equation}
\label{x2dCSL}
\begin{split}
&\frac{d \langle \hat{\ve{x}}^2 \rangle_t}{dt} = \frac{1}{m} \langle \hat{\ve{x}}\cdot\hat{\ve{p}} + \hat{\ve{p}}\cdot\hat{\ve{x}} \rangle_t -  \frac{\lambda A^{2}r_{C}^{3}}{2(\sqrt{\pi}\hbar)^{3}}\\
&\times \int d^3 Q \, \textrm{Tr} \left\{ \hat{\rho} [[\hat{\ve{x}}^2,L(\ve{Q},\hat{\ve{p}})], L(\ve{Q}, \hat{\ve{p}})] \right\}.
\end{split}
\end{equation}
After a long but straightforward calculation, one finds that
\begin{equation}
\label{x2dCSL1}
\begin{split}
& [[\hat{\ve{x}}^2,L(\ve{Q},\hat{\ve{p}})], L(\ve{Q}, \hat{\ve{p}})] = \\
&-\frac{8k^{2}r_{C}^{4}}{\hbar^{2}}\left[(1+k)\ve{Q}_{j}+2k\hat{\ve{p}}_{j}\right]^{2}L^{2}(\ve{Q},\hat{\ve{p}}).
\end{split}
\end{equation}
Using Eq.~\eqref{x2dCSL1} in Eq.~\eqref{x2dCSL}, and performing the trace over the momentum eigenvectors, the following integration appears:
\begin{equation}
\label{x2dCSL2}
\begin{split}
&\int d^{3}Q\,\int d^{3}p\,\left[(1+k)\ve{Q}+2k\ve{p}\right]^{2}e^{-\frac{r_{C}^{2}}{\hbar^{2}}\left[(1+k)\ve{Q}+2k\ve{p}\right]^{2}}\times \\
&\times \hat{\rho}(p,p,t)=\frac{3}{2}\left(\frac{\sqrt{\pi}}{1+k}\right)^{3}\left(\frac{\hbar}{r_{C}}\right)^{5} \\
\end{split}
\end{equation}
Collecting all results, we get:
\begin{equation}
\label{x2dCSL3}
\frac{d \langle \hat{\ve{x}}^2 \rangle_t}{dt} = \frac{1}{m} \langle \hat{\ve{x}}\cdot\hat{\ve{p}} + \hat{\ve{p}}\cdot\hat{\ve{x}} \rangle_t +\frac{6\lambda A^{2}r_{C}^{2}k^{2}}{\left(1+k\right)^{3}}.
\end{equation}
Note that for $k \to 0$ Eq.~\eqref{x2manyder} is recovered.
In order to solve Eq.~\eqref{x2dCSL3} we need to find $ \langle \hat{\ve{x}}\cdot\hat{\ve{p}} + \hat{\ve{p}}\cdot\hat{\ve{x}} \rangle_t$. The equation for $\langle \hat{\ve{x}}\cdot\hat{\ve{p}}  \rangle_t$ is: 
\begin{equation}
\label{xpdcsl}
\begin{split}
&\frac{d \langle \hat{\ve{x}} \cdot\hat{\ve{p}} \rangle_t}{dt} = \frac{1}{m} \langle \hat{\ve{p}}^2 \rangle_t -m\omega^{2}\left\langle \hat{\ve{x}}^{2}\right\rangle _{t}+ \frac{\lambda A^{2}r_{C}^{3}}{(\sqrt{\pi}\hbar)^{3}}\\
&\int d^3 Q \,\Bigl( \textrm{Tr} \left\{ e^{\frac{i}{\hbar} \ve{Q} \cdot \hat{\ve{x}}} L(\ve{Q}, \hat{\ve{p}}) \hat{\rho} (t) L(\ve{Q}, \hat{\ve{p}}) e^{-\frac{i}{\hbar} \ve{Q} \cdot \hat{\ve{x}}}  \hat{\ve{x}}\cdot \hat{\ve{p}}\right\}\\ 
& -\frac{1}{2} \textrm{Tr} \left\{ \left \{  L^2(\ve{Q}, \hat{\ve{p}}),  \hat{\rho} \right \} \hat{\ve{x}}\cdot \hat{\ve{p}}\right\} \Bigr)
\end{split}
\end{equation}
Using the ciclycity of the trace together with
\begin{equation}
e^{-\frac{i}{\hbar}\ve{Q} \cdot \hat{\ve{x}}}\,\hat{\ve{x}}\cdot\hat{\ve{p}}\,e^{\frac{i}{\hbar}\ve{Q} \cdot \hat{\ve{x}}}=\hat{\ve{x}}\cdot(\hat{\ve{p}}+\ve{Q})
\end{equation}
we can rewrite the trace as
\begin{equation}
\begin{split}
& \textrm{Tr} \left\{ e^{\frac{i}{\hbar} \ve{Q} \cdot \hat{\ve{x}}} L(\ve{Q}, \hat{\ve{p}}) \hat{\rho} (t) L(\ve{Q}, \hat{\ve{p}}) e^{-\frac{i}{\hbar} \ve{Q} \cdot \hat{\ve{x}}}  \hat{\ve{x}} \cdot \hat{\ve{p}}\right\}-\\
&-\frac{1}{2} \textrm{Tr} \left\{ \left \{  L^2(\ve{Q}, \hat{\ve{p}}),  \hat{\rho} \right \} \hat{\ve{x}} \cdot \hat{\ve{p}}\right\} =\\
&=\textrm{Tr}\left\{\hat{\rho}(t)L(\ve{Q},\hat{\ve{p}})\hat{\ve{x}}\cdot\ve{Q}L(\ve{Q},\hat{\ve{p}})\right\}-\\
&-\frac{1}{2}\textrm{Tr}\left\{\hat{\rho}(t)\left[\left[\hat{\ve{x}}\cdot\hat{\ve{p}},L(\ve{Q},\hat{\ve{p}})\right],L(\ve{Q},\hat{\ve{p}})\right]\right\}
\end{split}
\end{equation}
The term in the last line gives no contribution since the double commutator is zero. The integration over $\ve{Q}$ of the other term can be rewritten as follows:
\begin{equation}
\begin{split}
&\int d^{3}Q\,\textrm{Tr}\left\{\hat{\rho}(t)L(\ve{Q},\hat{\ve{p}})\hat{\ve{x}}\cdot\ve{Q}L(\ve{Q},\hat{\ve{p}})\right\}= \\
& =\frac{1}{2}\int d^{3}Q\,\textrm{Tr}\left\{L^{2}(\ve{Q},\hat{\ve{p}})\hat{\rho}\hat{\ve{x}}\cdot\ve{Q}\right\}\\
& +\frac{1}{2}\int d^{3}Q\,\textrm{Tr}\left\{L^{2}(\ve{Q},\hat{\ve{p}})\hat{\ve{x}}\cdot\ve{Q}\hat{\rho}\right\},
\end{split}
\end{equation}
%\vspace{-1 cm}
\\
and expanding the trace over the momentum eigenstates we get
\\
\\
\\
\begin{equation}
\begin{split}
\nonumber
&=\frac{1}{2}\sum_{j=1}^{3}\int d^{3}p\left(\int d^{3}QL^{2}(\ve{Q},\ve{p})Q_{j}\right)\langle\ve{p}|\left(\hat{\rho}\hat{x}_{j}\right)|\ve{p}\rangle \\
&+\frac{1}{2}\sum_{j=1}^{3}\int d^{3}p\left(\int d^{3}QL^{2}(\ve{Q},\ve{p})Q_{j}\right)\langle\ve{p}|\left(\hat{x}_{j}\hat{\rho}\right)|\ve{p}\rangle.
\end{split}
\end{equation}
Considering that
\begin{equation}
\int d^{3}QL^{2}(\ve{Q},\ve{p})Q_{j}=-\frac{2kp_{j}}{\left(1+k\right)}\left(\frac{\hbar\sqrt{\pi}}{\left(1+k\right)r_{C}}\right)^{3},
\end{equation}
\\
the dCSL contribution to Eq.~\eqref{xpdcsl} is
\begin{equation}
\begin{split}
&\frac{\lambda A^{2}r_{C}^{3}}{2(\sqrt{\pi}\hbar)^{3}}\int d^{3}Q\,\textrm{Tr}\Bigl(\hat{\rho}(t)L(\ve{Q},\hat{\ve{p}})\hat{\ve{x}}\cdot\ve{Q}L(\ve{Q},\hat{\ve{p}})\Bigr)=\\
&=-\frac{\lambda A^{2}k}{\left(1+k\right)^{4}}\left\langle \hat{\ve{x}}\cdot\hat{\ve{p}}+\hat{\ve{p}}\cdot\hat{\ve{x}}\right\rangle _{t}.
\end{split}
\end{equation}
So, the equation for $\left\langle \hat{\ve{x}}\cdot\hat{\ve{p}}\right\rangle _{t}$ is:
\begin{equation}
\frac{d\left\langle \hat{\ve{x}}\cdot\hat{\ve{p}}\right\rangle _{t}}{dt}=\frac{1}{m}\left\langle \hat{\ve{p}}^{2}\right\rangle _{t}-m\omega^{2}\left\langle \hat{\ve{x}}^{2}\right\rangle _{t}-\frac{\lambda A^{2}k}{\left(1+k\right)^{4}}\left\langle \hat{\ve{x}}\cdot\hat{\ve{p}}+\hat{\ve{p}}\cdot\hat{\ve{x}}\right\rangle _{t}
\end{equation}
which implies that:
\begin{equation}
\label{xppxdcsl}
\begin{split}
&\frac{d\left\langle \hat{\ve{x}}\cdot\hat{\ve{p}}+\hat{\ve{p}}\cdot\hat{\ve{x}}\right\rangle _{t}}{dt}=\frac{2}{m}\left\langle \hat{\ve{p}}^{2}\right\rangle _{t}-2m\omega^{2}\left\langle \hat{\ve{x}}^{2}\right\rangle _{t} \\
&-\frac{2\lambda A^{2}k}{\left(1+k\right)^{4}}\left\langle \hat{\ve{x}}\cdot\hat{\ve{p}}+\hat{\ve{p}}\cdot\hat{\ve{x}}\right\rangle _{t}.
\end{split}
\end{equation}
%In order to solve Eq.~\eqref{xppx} we substitute the first term in the right hand side of the equation using Eq.~\eqref{energydCSL} and defining 
%\begin{equation}
%\label{cBD}
%\begin{split}
%& c(t):=\left\langle \hat{\ve{x}}\cdot\hat{\ve{p}}+\hat{\ve{p}}\cdot\hat{\ve{x}}\right\rangle _{t}, \\
%& B:=\frac{2\lambda A^{2}m^{2}k}{\left(1+k\right)^{4}m_{0}^{2}}=\frac{\left(1+k\right)}{2}\chi, \\
%& D:=H(0)-H_{as}.
%\end{split}
%\end{equation}
%Then Eq.~\eqref{xppx} becomes
%\begin{equation}
%\frac{dc(t)}{dt}=4De^{-\chi t}+4H_{as}-Bc(t).
%\end{equation}
%The solution of this equation is:
%\begin{eqnarray}
%c(t)&=&\frac{4D}{\left(B-\chi\right)}e^{-\chi t}+\left(c(0)-\frac{4D}{\left(B-\chi\right)}-\frac{4H_{as}}{B}\right)e^{-Bt}\nonumber\\
%\nonumber\\
%&+&\frac{4H_{as}}{B}.
%\end{eqnarray}

The last equation we need is that for the momentum variance $\left\langle \hat{\ve{p}}^{2}\right\rangle _{t}$. This has been already derived in~\cite{andrea}:
\begin{equation}
\label{dCSLp2}
\frac{d\left\langle \hat{\ve{p}}^{2}\right\rangle _{t}}{dt} = -m\omega^2 \left\langle \hat{\ve{x}}\cdot\hat{\ve{p}}+\hat{\ve{p}}\cdot\hat{\ve{x}}\right\rangle _{t} - \chi\left\langle \hat{\ve{p}}^{2}\right\rangle _{t} + \chi \left\langle \hat{\ve{p}}^{2}\right\rangle _{as},
\end{equation}
where
\begin{equation}\label{chidCSL}
\chi := \frac{4k\lambda A^2}{(1+k)^5} , \qquad \left\langle \hat{\ve{p}}^{2}\right\rangle _{as}:= \frac{3\hbar^2}{8 k r_{C}^{2}}. 
\end{equation}
%
% and 
% \begin{equation}\label{has}
% \left\langle \hat{\ve{p}}^{2}\right\rangle _{as}= \frac{3\hbar^2}{8 k r_{C}^{2}}. 
% \end{equation}
%
In the limit of free evolution (\textit{i.e.} $\omega \to 0$), the solutions of  Eqs.~\eqref{x2dCSL3}, \eqref{xppxdcsl} and \eqref{dCSLp2} are:
\begin{widetext}
\begin{subequations}
\label{dcslfree}
\begin{align}
&\left\langle \hat{\ve{x}}^{2}\right\rangle _{t} = \left\langle \hat{\ve{x}}^{2}\right\rangle _{t_0} +\frac{ 2(\left\langle \hat{\ve{p}}^{2}\right\rangle _{t_0} - \left\langle \hat{\ve{p}}^{2}\right\rangle _{as})}{m^2 (B-\chi)} \left(\frac{1-e^{-\chi (t-t_0)}}{\chi} - \frac{1-e^{-B(t-t_0)}}{B}\right) +  \left(  \left\langle \hat{\ve{x}}\cdot\hat{\ve{p}}+\hat{\ve{p}}\cdot\hat{\ve{x}}\right\rangle _{t_0} - \frac{2 \left\langle \hat{\ve{p}}^{2}\right\rangle _{as}}{m B} \right) \times \notag \\ 
&\times \frac{1-e^{-B(t-t_0)}}{m B}+ \left(\alpha+  \frac{2 \left\langle \hat{\ve{p}}^{2}\right\rangle _{as}}{m^2 B}\right)(t-t_0) \label{x2dcslfree}; \\ 
&\left\langle \hat{\ve{x}}\cdot\hat{\ve{p}}+\hat{\ve{p}}\cdot\hat{\ve{x}}\right\rangle _{t} = \frac{2 ( \left\langle \hat{\ve{p}}^{2}\right\rangle_{t_0} - \left\langle \hat{\ve{p}}^{2}\right\rangle_{as})}{m(B-\chi)} \left (e^{-\chi (t-t_0)} - e^{-B(t-t_0)} \right) +  \frac{2 m\left\langle \hat{\ve{p}}^{2}\right\rangle_{as}}{m B} + e^{-B(t-t_0)}\times \notag \\
&\times \left( \left\langle \hat{\ve{x}}\cdot\hat{\ve{p}}+\hat{\ve{p}}\cdot\hat{\ve{x}}\right\rangle_{t_0}- \frac{8 m \left\langle \hat{\ve{p}}^{2}\right\rangle_{as}}{B} \right) \label{xppxdcslfree}; \\
&\left\langle \hat{\ve{p}}^{2}\right\rangle_{t} = \left\langle \hat{\ve{p}}^{2}\right\rangle_{as} + e^{-\chi (t-t_0)}\left(\left\langle \hat{\ve{p}}^{2}\right\rangle_{t_0} -\left\langle \hat{\ve{p}}^{2}\right\rangle_{as}\right) \label{p2dcslfree},
\end{align}
\end{subequations}
\end{widetext}
where $B := \frac{1+k}{2} \chi$ and $\alpha := \frac{6\lambda A^{2}r_{C}^{2}k^{2}}{\left(1+k\right)^{3}}$. 

We should study  also the case of an harmonically trapped atom ($\omega \neq 0$). The system of Eqs.~\eqref{x2dCSL3}, \eqref{xppxdcsl} and \eqref{dCSLp2} can still be solved exactly. However, the solutions are too complicated and of little practical use. In fact, the duration of the delta-kick is much shorter than the free evolution and, as  shown in the Appendix,  the dCSL effects during the delta-kick can be neglected and safely be replaced by the standard quantum mechanical evolution. 

We can now derive the position variance $\langle \hat{\ve{x}}^2 \rangle_{t_3}$ at the final time $t_3$ as predicted by the dCSL model. During  steps 1 and 3 of the experiment (free expansion of the gas) we use the exact solutions given in Eqs.~\eqref{x2dcslfree}, \eqref{xppxdcslfree} and \eqref{p2dcslfree}, while during step 2 (the delta-kick) we use the quantum mechanical solution for an harmonic oscillator. In a similar way, one can compute the time evolution of the average kinetic energy.  We do not report explicitly the final formula for $ \langle \hat{\ve{x}}^2 \rangle_{t_3}$ since it is very long and does not help in getting any insight on the physics. 
% We computed the position variance of the gas at the detector (time $t=t_2$). Due to the major complexity of the dCSL model, in this case the final result is quite complicated (see Appendices 3-4 for details) and in particular it cannot be split as a sum of a quantum-mechanical and a dCSL term.

\begin{figure}[t]
\includegraphics[scale=0.7]{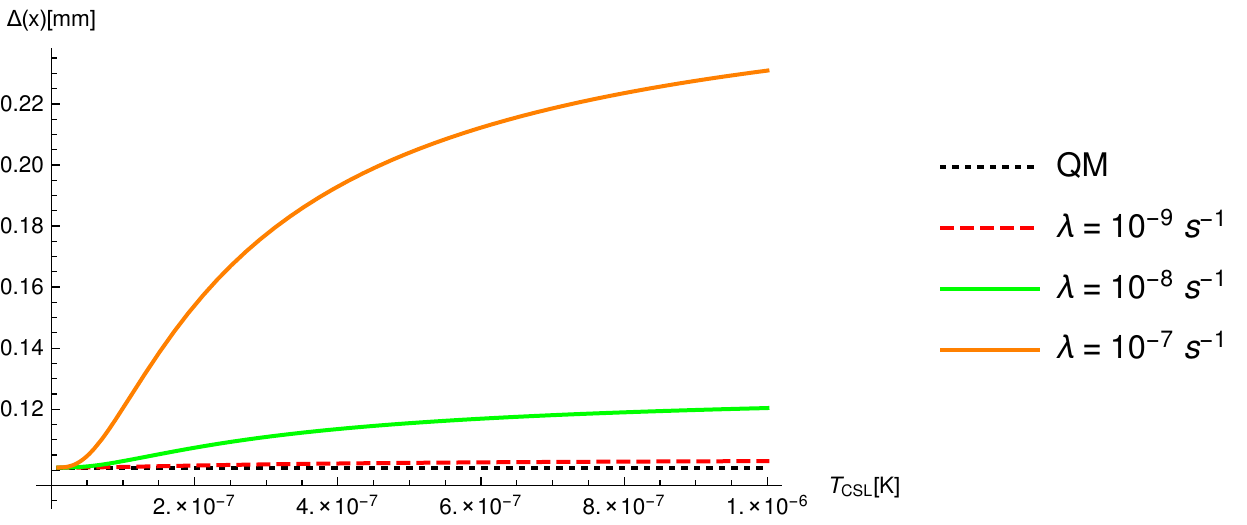}
\caption{Position's standard deviation $\Delta (x) \equiv \langle \hat{\ve{x}}^2 \rangle_{t_3}^{1/2}$ at the detector (time $t = t_3$) as a function of the CSL noise temperature $T_{\text{\tiny{CSL}}}$, for three different values of the collapse rate $\lambda$. For each curve, we fixed $r_C = 10^{-7}$ m and the delta-kick time $\delta t_2  = 35$ ms, which corresponds to the smallest measured value (black point in Fig.~\ref{pos-taup}). We plot also the quantum-mechanical value  for comparison. }
\label{pos-temp} 
\end{figure}

\begin{figure}[t]
\includegraphics[scale=0.7]{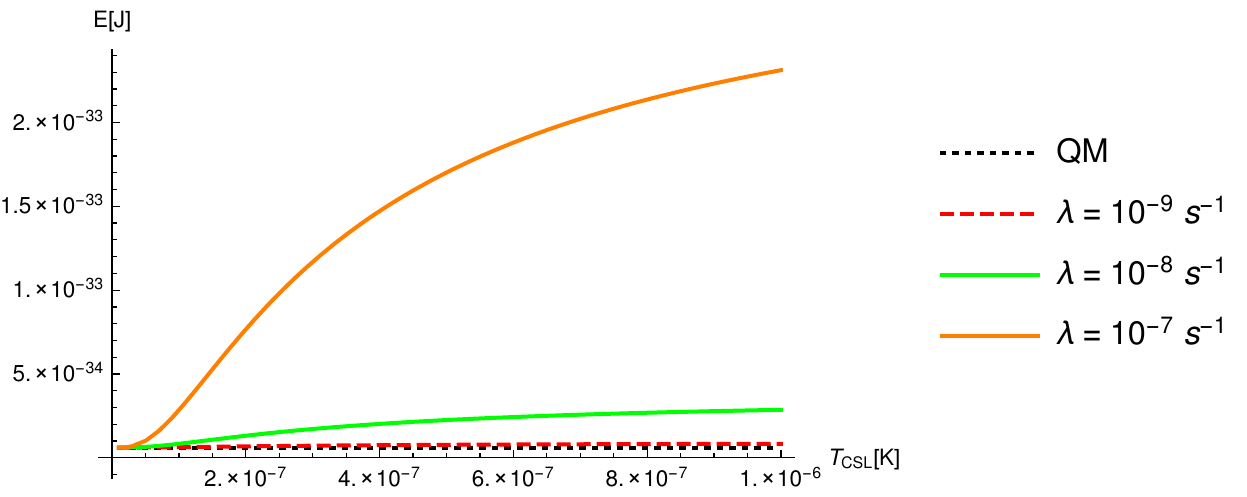}
\caption{Kinetic energy $E \equiv \langle \hat{\ve{p}}^2 \rangle_{t_3}/ 2m $ at the detector (time $t = t_3$) as a function of the CSL noise temperature $T_{\text{\tiny{CSL}}}$, for three different values of the collapse rate $\lambda$. For each curve, we fixed $r_C = 10^{-7}$ m and $\delta t_2 = 20$ ms, associated to the theoretical minimum of the momentum standard deviation (see Fig.~\ref{kin-taup}). We plot also the quantum-mechanical value for comparison.}
\label{kin-temp}
\end{figure}

In Fig.~\ref{pos-temp} and Fig.~\ref{kin-temp} the minimum values of the final position variance and of the average kinetic energy are plotted as a function of the noise temperature $T_{\text{\tiny{CSL}}}$ for different values of $\lambda$, while keeping $r_C=10^{-7}$m, and for fixed values of the delta-kick time (we took the values of $\delta t_2$ which maximize the delta-kick effects). We can see that in both cases the effect of dissipation is to reduce the increase of the variance and of the energy due to the CSL noise. In particular, for the values of $r_C$ and $\lambda$ here considered, when $T_{\text{\tiny{CSL}}}< 10^{-7}$ K the effect of the noise is negligible and the predictions are practically equivalent to the standard quantum ones. In the range $10^{-7} \textrm{ K}<T_{\text{\tiny{CSL}}}< 10^{-6} \textrm{ K}$ the noise effects are present but  are reduced by  dissipation.  When $T_{\text{\tiny{CSL}}}> 10^{-6} \textrm{ K}$ the effects of dissipation become negligible and the predictions are indistinguishable from the $T_{\text{\tiny{CSL}}}=+\infty$ case (CSL).  
%
%%%%%%%%%%%%%%%%%%%%%%%%%%%%%%%%%%%%%%%%%%%%%%%%%%%%%%%%%%%%%%%%%%%%%
\subsection{dCSL model with boost}
%%%%%%%%%%%%%%%%%%%%%%%%%%%%%%%%%%%%%%%%%%%%%%%%%%%%%%%%%%%%%%%%%%%%%%
The dCSL model is not Galilei invariant, since the noise selects a preferred reference frame, the one where it is at rest. In the previous section, we implicitly considered the situation where the lab reference frame was at rest with respect to the noise. This is unlikely. If the noise has a cosmological origin,  then much likely it is at rest with the cosmic frame, with respect to which the Earth moves. In this section we analyse the case where the collapse noise is moving with some velocity $\ve{u}$ with respect to the laboratory system. 
% One reason for considering this model is given by the hypothesis that the collapse noise could have a cosmological origin. In such a case it is reasonable to suppose this noise being at rest with respect to the fixed stars inertial frame. However the same noise is not seen at rest in a laboratory in the Earth, since the Earth moves with respect to the fixed stars inertial frame.

The master equation for the boosted dCSL model has the same structure as that in Eq.~\eqref{dCSL} with $L(\ve{Q}, \hat{\ve{p}})$ in Eq.~\eqref{LdCSL} replaced by:
\begin{equation}\label{LdCSLbst}
L(\ve{Q},\hat{\ve{p}},\ve{u}) =  e^{-\frac{r_{C}^{2}}{2\hbar^2} \left \lvert(1+k)\ve{Q} + 2k(\hat{\ve{p}}-m\ve{u}) \right\rvert^2}.
\end{equation}
It is convenient to introduce the boosted momentum operator:
\begin{equation}\label{p_bst}
\hat{\ve{p}}_{\ve{u}}:=\hat{\ve{p}}-m\ve{u}
\end{equation}
which allows to rewrite the boosted dCSL  master equation as
\begin{equation}\label{masterboost}
\frac{d\hat{\rho}}{dt} = \left.\frac{d\hat{\rho}^{\text{\tiny dCSL}}}{dt}\right|_{\hat{\ve{p}}
\rightarrow\hat{\ve{p}}_{\ve{u}}}-\frac{i}{\hbar}\left[\hat{\ve{p}}_{\ve{u}}\cdot\ve{u},\hat{\rho}\right]
\end{equation}
where the first term is the master equation of the dCSL as given by Eq.~\eqref{dCSL}, with  $\hat{\ve{p}}_{\ve{u}}$ in place of $\hat{\ve{p}}$. Note that $\hat{\ve{p}}_{\ve{u}}$ has the same commutation relations as $\ve p$. The equation for the time evolution of a generic operator $O$ is:
\begin{equation}\label{operatorboost}
\frac{d\left\langle O\right\rangle _{t}}{dt}= \left.\frac{d\left\langle O\right\rangle _{t}^{\text{\tiny dCSL}}}{dt}\right|_{\hat{\ve{p}}
\rightarrow\hat{\ve{p}}_{\ve{u}}}-\frac{i}{\hbar}\left\langle \left[O,\hat{\ve{p}}_{\ve{u}}\cdot\ve{u}\right]\right\rangle _{t}
\end{equation}
where $\left\langle O\right\rangle^{\text{\tiny dCSL}}_{t}$ is the expectation value of the operator $O$ given by the dCSL dynamics without boost, again with $\hat{\ve{p}}_{\ve{u}}$ in place of $\hat{\ve{p}}$. Therefore, we can now write the equations for the expectation values $\langle \hat{\ve{x}}^2 \rangle _{t}$, $\langle \hat{\ve{p}}_{\ve{u}}^2 \rangle _{t}$ and $\left\langle \hat{\ve{x}}\cdot \hat{\ve{p}}_{\ve{u}}+\hat{\ve{p}}_{\ve{u}}\cdot\hat{\ve{x}}\right\rangle_{t}$ using the results already derived for the dCSL model without boost; we only need to compute the extra commutator of Eq.~\eqref{operatorboost}.

\begin{figure}[t]
\includegraphics[scale=0.4]{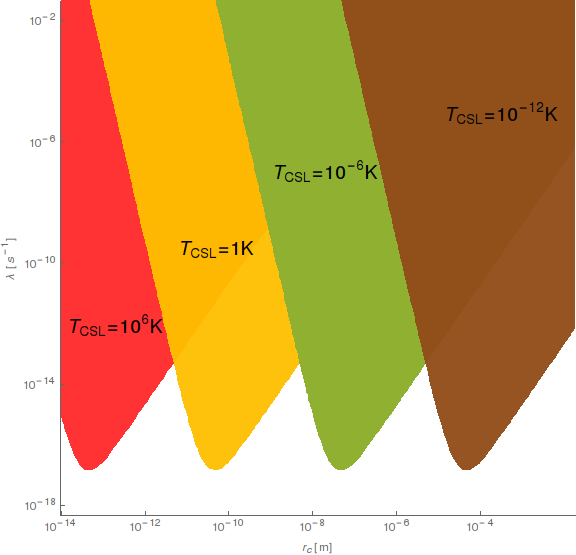}
\caption{Exclusion plot  for the boosted dCSL model, considering a boost with $\left \lvert \ve{u} \right \rvert = 10^7$ m$\textit{s}^{-1}$ for four different values of the dCSL temperature $T_{\text{\tiny{CSL}}}$.}
\label{exc_boost}
\end{figure}

Actually,  to get a good estimate of the effect of the boost, it is sufficient to analyze  the equations for $\langle \hat{\ve{x}} \rangle _{t}$ and $\langle \hat{\ve{p}} \rangle _{t}$, instead of those for the variances, which are much more complicated. The first equation can be easily derived, while the second one  involves lengthier calculations, which however are analogue to those carried out in the previous section, when deriving the equation for $\langle \hat{\ve{x}} \cdot\hat{\ve{p}} \rangle_t$. The final result is:
\begin{equation}\label{x_u}
\frac{d\left\langle \hat{\ve{x}}\right\rangle _{t}}{dt}=\frac{\left\langle \hat{\ve{p}}_{\ve{u}}\right\rangle _{t}}{m}+\ve{u},\;\;\;\;\;\;
\frac{d\left\langle \hat{\ve{p}}_{\ve{u}}\right\rangle _{t}}{dt}=-B\left\langle \hat{\ve{p}}_{\ve{u}}\right\rangle _{t},
\end{equation}
where $B$ is the parameter defined after Eq.~\eqref{p2dcslfree}.
%
%The equations for the variances and correlations are:
%\begin{equation}\label{x2_u}
%\frac{d\left\langle \hat{\ve{x}}^{2}\right\rangle _{t}}{dt}=\frac{1}{m}\left\langle \hat{\ve{x}}\cdot\hat{\ve{p}}_{\ve{u}}+\hat{\ve{p}}_{\ve{u}}\cdot\hat{\ve{x}}\right\rangle _{t}+\frac{6\lambda r_{C}^{2}m^{2}k^{2}}{m_{0}^{2}\left(1+k\right)^{3}}+2\ve{u}\cdot\left\langle \hat{\ve{x}}\right\rangle _{t};
%\end{equation}
%\begin{eqnarray}\label{xp+px_u}
%&&\frac{d\left\langle \hat{\ve{x}}\cdot\hat{\ve{p}}_{\ve{u}}+\hat{\ve{p}}_{\ve{u}}\cdot\hat{\ve{x}}\right\rangle _{t}}{dt}=\frac{2}{m}\left\langle \hat{\ve{p}}_{\ve{u}}^{2}\right\rangle _{t}-2m\omega^{2}\left\langle \hat{\ve{x}}^{2}\right\rangle _{t}\nonumber\\
%\nonumber\\
%&&-\frac{2\lambda A^{2}m^{2}k}{\left(1+k\right)^{4}m_{0}^{2}}\left\langle \hat{\ve{x}}\cdot\hat{\ve{p}}_{\ve{u}}+\hat{\ve{p}}_{\ve{u}}\cdot\hat{\ve{x}}\right\rangle _{t}+2\ve{u}\cdot\left\langle \hat{\ve{p}}_{\ve{u}}\right\rangle _{t};
%\end{eqnarray}
%\begin{equation}\label{p2_u}
%\frac{d\left\langle \hat{\ve{p}}_{\ve{u}}^{2}\right\rangle _{t}}{dt}=-m\omega^{2}\left\langle \hat{\ve{x}}\cdot\hat{\ve{p}}_{\ve{u}}+\hat{\ve{p}}_{\ve{u}}\cdot\hat{\ve{x}}\right\rangle _{t}-\chi\left\langle \hat{\ve{p}}_{\ve{u}}^{2}\right\rangle _{t}+\chi\left\langle p^{2}\right\rangle _{as}.
%\end{equation}
%
The solution of this system of equations for a free gas ($\omega=0$) with initial average position $\langle \hat{\ve{x}} \rangle _{t_0}$ and initial average momentum $\langle \hat{\ve{p}} \rangle _{t_0}$, written in terms of the real momentum $\hat{\ve{p}}$, are:
%\begin{widetext}
\begin{equation}\label{x_usol}
\left\langle \hat{\ve{x}}\right\rangle _{t} = \left\langle \hat{\ve{x}}\right\rangle _{t_0}  + \ve{u}(t-t_0) + \left( \frac{\left\langle \hat{\ve{p}}\right\rangle _{t_0}}{m} - \ve{u} \right) \frac{1-e^{-B(t-t_0)}}{B};
\end{equation}
\begin{equation}\label{p_usol}
\left\langle \hat{\ve{p}}\right\rangle _{t} = \left\langle \hat{\ve{p}}\right\rangle _{t_0}e^{-B(t-t_0)} + m\ve{u}\left( 1-e^{-B(t-t_0)}\right);
\end{equation}
We can now argue as follows. The change of the average position of the gas must be smaller than the measured standard deviation, as in~\cite{kas} no significant variation to the average position of the center-of-mass of the cloud was observed. From Eq.~\eqref{x_usol}, taking into account that for the experiment considered here $\left\langle \hat{\ve{p}}\right\rangle _{t_0} = 0$ and $t-t_0 \approx 3$ s, and that for any value of the parameters of the dCSL model $B (t-t_0) \ll 1$, we can safely say that
\begin{equation}
\label{ubound}
\frac{1}{2} \left \lvert \ve{u} \right \rvert B (t-t_0)^2\leq 1 \, \mu\text{m}.
\end{equation}
where $B = 2 \lambda A^2 k/(1+k)^4$. Considering, for example, the standard values for the dCSL parameters $\lambda=10^{-17}$ s$^{-1}$, $r_C=10^{-7}$ m and $T_{\text{\tiny{CSL}}}=1$ K, we obtain the bound:
\begin{equation}
\left \lvert \ve{u} \right \rvert \leq 10^{13} \, \text{m} \, \text{s}^{-1}.
\end{equation}
From cosmological arguments~\cite{arbest} a possible value of the noise boost is $\left \lvert \ve{u} \right \rvert = 10^7$ m$\textit{s}^{-1}$. Using this value in Eq.~\eqref{ubound} an exclusion plot in the parametric space $\lambda - r_C$ is found, as shown in Fig.~\ref{exc_boost}.
%
%%%%%%%%%%%%%%%%%%%%%%%%%%%%%%%%%%%%%%%%%%%%%%%%%%%%%%%%%%%%%%%%%%%%%
\section{Discussion: comparison with experimental data and bounds on the collapse parameters}
\begin{figure}[t]
\includegraphics[scale=0.4]{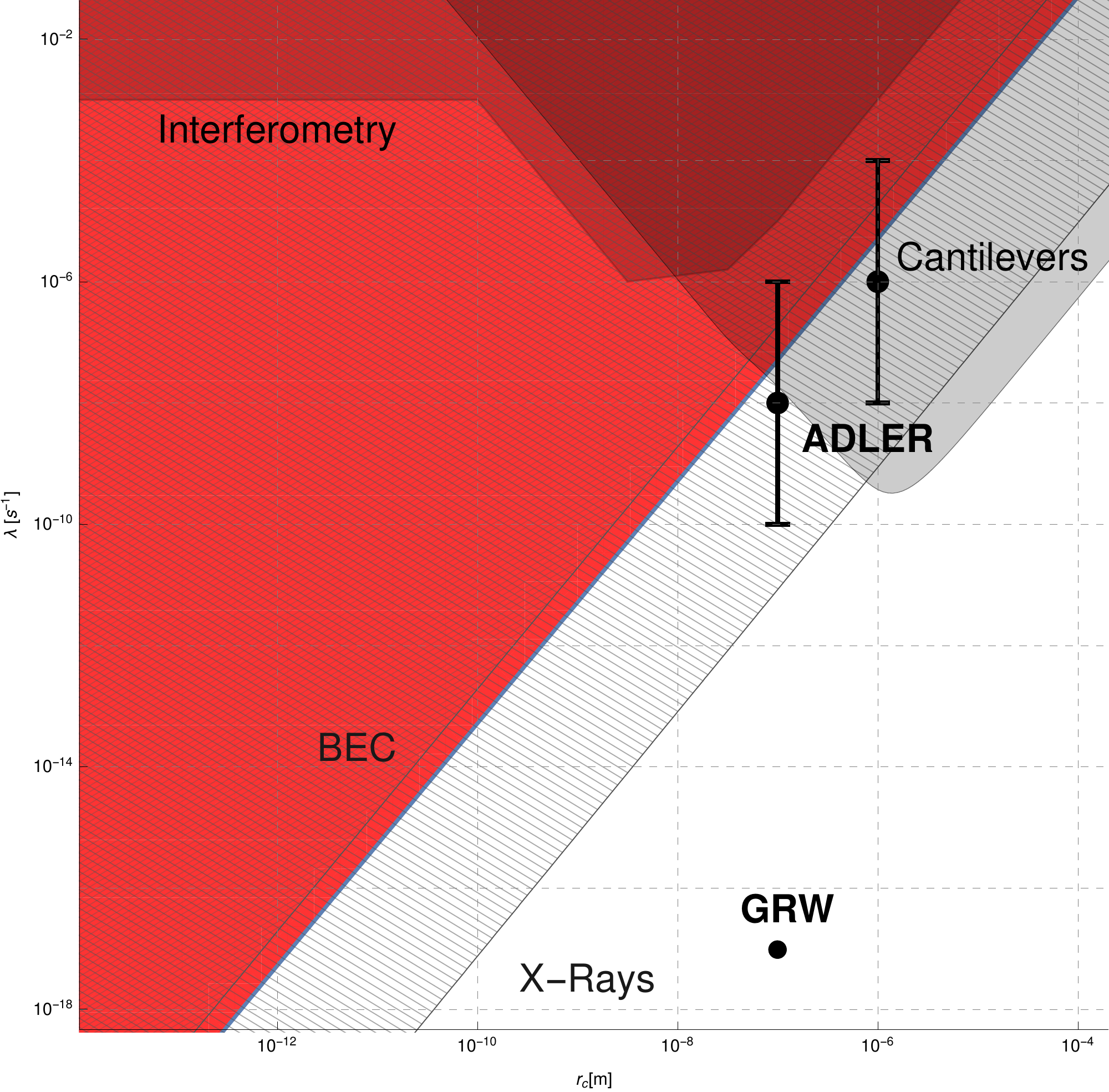}
\caption{Exclusion plot for the CSL model. The red region shows the excluded area, according to the analysis here performed. The picture shows also the bounds coming from matter-wave interferometry~\cite{arbest},  cantilevers~\cite{vin}, heating effect on Bose-Einstein Condensates (BECs)~\cite{pearle}, and spontaneous X-rays emission~\cite{bea}. The black points and bars represents the reference values proposed by GRW ~\cite{grw} and Adler~\cite{ad1}.}
\label{exc_csl}
\end{figure}

\begin{figure}[t]
\includegraphics[scale=0.4]{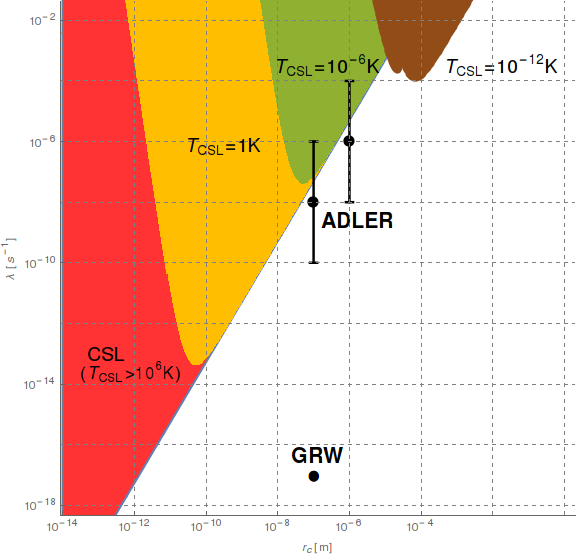}
\caption{Exclusion plot for the dCSL model. The red area represents the excluded region for the CSL model ($T = \infty$) and for any dCSL model with noise temperatures $T_{\text{\tiny{CSL}}} > 10^6$ K (due to the finite parametric region considered). Bounds for dCSL for three different noise temperatures are also represented: in yellow the case with $T_{\text{\tiny{CSL}}} = 1$ K, in green that for $T_{\text{\tiny{CSL}}} = 10^{-6}$ K, and  in brown $T_{\text{\tiny{CSL}}} = 10^{-12}$ K. The black points and bars represents the parametric values proposed by GRW ~\cite{grw} and Adler~\cite{ad1}.}
\label{exc_dcsl}
\end{figure}

%%%%%%%%%%%%%%%%%%%%%%%%%%%%%%%%%%%%%%%%%%%%%%%%%%%%%%%%%%%%%%%%%%%%%%
We now discuss the bounds on the collapse parameters against the experiment here considered. We compare the position's standard deviation, computed for each particular model, with the experimental value $\Delta (x)_{\text{\tiny{EXP}}} = 120_{-40}^{+40} \, \mu$m reported in~\cite{kas}; we refer to this value since it is the only one with explicit error bars associated to it. 
Assuming that this value is distributed according to a normal distribution with mean value $ \mu=120\, \mu$m and $\sigma=40\, \mu$m, then $\Delta (x) \in [42;198] \, \mu$m with a confidence level of 95\%
%Note that the quantum mechanical prediction $\Delta (x)_{\text{\tiny{QM}}}=(\langle \hat{\ve{x}}^2 \rangle_{ t_3}^{\text{\tiny{QM}}})^{1/2}=100\, \mu$m (see Eq.~\eqref{posvart2qm}) falls within this range.
%Translated into variance, the above experimental value becomes $\Delta x^2_{\text{\tiny{EXP}}} = 1.6_{1}^{1} \times 10^4 \mu$m$^2$. According to Eq.~\eqref{posvart2}) the total variance is the sum of the CSL contribution and the standard quantum mechanics one. By taking the $2\sigma$ error bar and subtracting the latter term we find an experimental upper bound on the CSL contribution $\langle \hat{\ve{x}}^2 \rangle_{t_3}^{\text{\tiny{CSL}}}< 2.6 \times 10^4$ $\mu$m$^2$ with approximately 95\% confidence level \footnote{For a more rigorous analysis, we should also take into account that collapse models can only increase the position's standard deviation of the cloud (with the exception of the dCSL model with $T_{\text{\tiny{CSL}}} \leq 10^{-12}$ K). This implies that values of $\Delta(x)$ lower than the quantum mechanical prediction ($\Delta(x)_{\text{\tiny QM}} = 100 \mu m$) are excluded a priori. This analysis can be done according to~\cite{feldman} and leads essentially to the same upper bound, so it does not modify the analysis of Section VI.}. 
The exclusion plots in Fig.~\ref{exc_csl} and Fig.~\ref{exc_dcsl} show which points in the parameters space predict a CSL-induced position's standard deviation outside the considered range (with a  $T_{\text{\tiny{CSL}}}$ dependence in the dCSL case).

We start with analysing the CSL model. As shown in Eq.~\eqref{posvart2csl}, the increase of the position variance at the final time $t=t_3$ due to the CSL noise is: 
\begin{equation}
\label{spreadcsl}
\langle \hat{\ve{x}}^2 \rangle_{t_3}^{\text{\tiny{CSL}}} = \frac{\lambda}{r_{C}^{2}} K
\end{equation}
where $K$ is a function of the initial state of the gas, the times $t_1, \delta t_2,  t_3$ and the frequency $\omega$ of the external harmonic potential, but otherwise contains no dependence on the CSL parameters. By inserting the numerical values, we arrive at the bound:
\begin{equation}
\frac{\lambda}{r_{C}^{2}} < 5 \times 10^6 \,\textrm{m}^{-2}\textrm{s}^{-1}.
\end{equation}
\begin{figure}[t]
\includegraphics[scale=0.72]{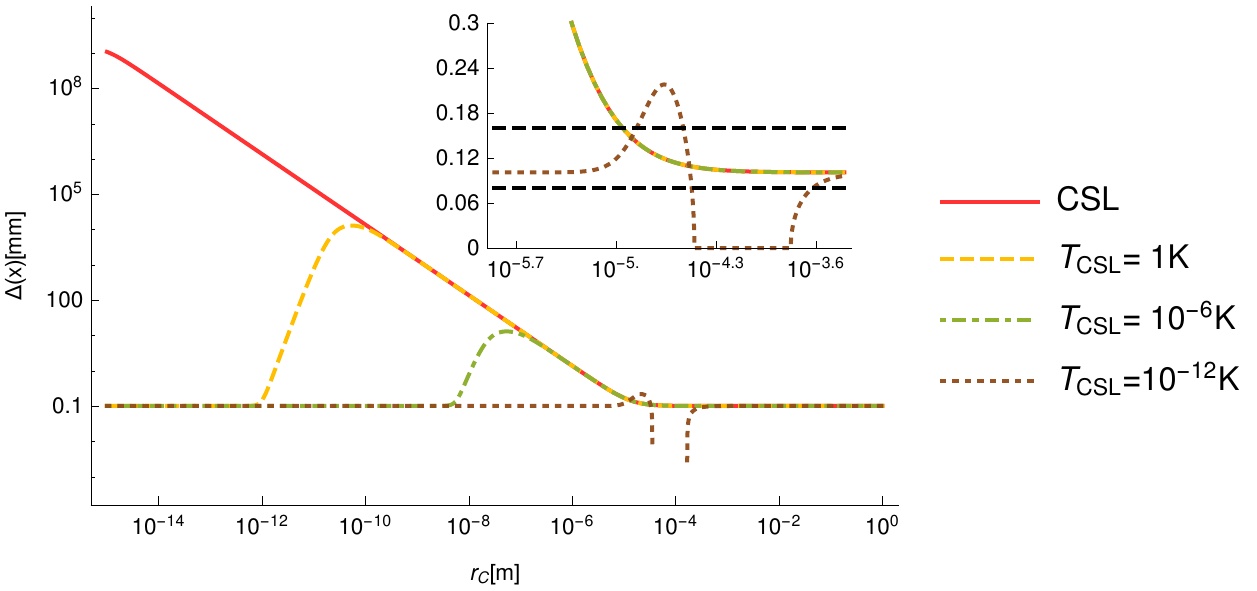}
\caption{Position's standard deviatio of the gas at the detector (time $t = t_3$), as a function of  $r_C$. Four different curves are represented, each corresponding to a different value of the noise temperature $T_{\text{\tiny{CSL}}}$. In each case, $\lambda = 10^{-3.5}$ s$^{-1}$ and $\delta t_2 = 35$ ms. For $T_{\text{\tiny{CSL}}} = 10^{-12}$ K, an insignificant numerical error appears in the interval $10^{-4.4}$ m $<r_C< 10^{-3.7}$ m, due to very small values of the position's standard deviation. No numerical instabilities appear outside this interval. The experimental value $120_{-40}^{+40} \,\mu$m is indicated by the dashed black lines in the inset.}
\label{devrc}
\end{figure}
This result is in agreement with the plot in Fig.~\ref{exc_csl}, where  a comparison with  bounds coming from other relevant experiments is shown. As one can see, the bound is better than that coming from matter-wave interferometry~\cite{arbest} and that related to BECs~\cite{pearle} while,  for $r_C\leq 10^{-7}$, it is beaten only by  X-rays experiments~\cite{bea}. Here a comment is at order. As shown in~\cite{adlerrama, dirk}, CSL predictions for spontaneous photon emission are very sensitive to the type of noise and, when a frequency cut-off is introduced  in its spectrum, the CSL effect is significantly decreased. In particular, for X-ray detection, any cutoff smaller than $10^{18}$ Hz washes the effect away. Since typical cut-offs of cosmological spectra are significantly smaller than $10^{18}$ Hz~\cite{bas}, and assuming that the CSL noise has the properties of a typical cosmological random background, then one expects bounds related to spontaneous X-ray emission not to play a significant role. On the other hand, our result is robust against changes in the noise. As shown in Sec.~\ref{color}, providing $r_C \geq 10^{-7}$ m, for any cutoff larger than $10^6$ Hz (which is the case of cosmological noises), the effect is equivalent to that of the standard CSL model.

The situation si different for the dCSL model. The result is reported in Fig.~\ref{exc_dcsl}, for three different temperatures of the noise: $T_{\text{\tiny{CSL}}}= 1, 10^{-6}, 10^{-12}$ K.  As one can see, the smaller the temperature, the smaller the exclusion region. The reason is that dissipation reduces the Brownian motion fluctuations of the atoms, therefore also the extra spread of the position variance predicted by CSL. 
% For $T_{\text{\tiny{CSL}}}=1 and 10^{-6}$K, the predictions of dCSL are practically equivalent to the ones of the standard CSL model, except for really small values of $r_C$. This is a consequence of the dependence of the parameter $k$ defined in Eq.~\eqref{k} from $r_C$: given a fixed value for the noise temperature $T_{\text{\tiny{CSL}}}$, decreasing the value of $r_C$ is equivalent to send $k \to + \infty$, and quantum mechanical predictions are recovered. Viceversa, higher values of $r_C$ implies $k \to 0$ and CSL predictions are recovered. This can be seen clearly for noise temperature $T_{\text{\tiny{CSL}}}= 1, 10^{-6}$K in fig.~\ref{exc_dcsl}. 
%
The case $T_{\text{\tiny{CSL}}}= 10^{-12}$ is significant. In fact, a noise temperature of the order of 1 picokelvin is lower than the system's temperature, and the dissipative dynamics cools the system, reducing its position and momentum spread. For this reason, the excluded area in the parameter spaces it is fundamentally different from the other, high-temperature situations. 
% We use this results to build an exclusion plot for $\lambda$ and $r_C$, see fig.~\ref{exc_dcsl}. We note that, as the temperature of the noise $T_{\text{\tiny{CSL}}}$ decreases and the effects of dissipation becomes more relevant the parameter region excluded by the experiment reduces. 

Also the shape of the curve for $T_{\text{\tiny{CSL}}}=10^{-12}$ K is different from the other cases. This can be better seen in Fig.~\ref{devrc} where, for fixed $T_{\text{\tiny{CSL}}}$  and $\lambda=10^{-3.5}$ s$^{-1}$, the final position variance $\langle \hat{\ve{x}}^2 \rangle_{t_3}$ is plotted as function of $r_C$. 
% For $T_{\text{\tiny{CSL}}}= 1$K, $10^{-6}$ K, the noise acts as an heating source on the system, due to the temperature difference. This effect is similar to the CSL model, corresponding to $T_{\text{\tiny{CSL}}} \to +\infty$. Instead, for $T_{\text{\tiny{CSL}}}= 10^{-12}$ K, the noise cools the system, reducing its position variance. This is clearly shown in the inset of fig.~\ref{devrc}. The initial increasing of the position standard deviation, as $r_C$ increase, is due to an expansion of the gas in the dCSL model, intrinsic for every temperature. This is expressed by the linear term in ~eq \eqref{dcslfree}. 
%
% For any temperature,  we note the dependence of the position standard deviation as $r_C \to 0$ and $r_C \to +\infty$. From  eq.~\eqref{k}, we note that the limit $r_C \to 0$ corresponds to have $k \to +\infty$, giving the quantum mechanical predictions. On the other case, the limit $r_C \to +\infty$ corresponds to the CSL limit of $k \to 0$.

To conclude, the bounds on the CSL parameters coming from the experiment in~\cite{kas} are among the strongest so far analysed, stronger than direct tests based on matter-wave interferometry. They are robust against changes in the spectrum of the noise, so in this sense they are the strongest for $r_C < 10^{-7}$ m. They become weaker when dissipation is included, still remaining strong down to very small temperatures.

%
%%%%%%%%%%%%%%%%%%%%%%%%%%%%%%%%%%%%%%%%%%%%%%%%%%%%%%%%%%%%%%%%%%%%%
% \section{Conclusions}
%%%%%%%%%%%%%%%%%%%%%%%%%%%%%%%%%%%%%%%%%%%%%%%%%%%%%%%%%%%%%%%%%%%%%%
% Our results are summarized in fig.~\ref{exc_csl} and fig.~\ref{exc_dcsl}. The comparison with experimental data sets, for the CSL model, the bound $\lambda/r_C^2 \leq 10^{6}$ s$^{-1}$ m$^{-2}$ which implies, when $r_C=10^{-7}$ m, to set $\lambda \leq 10^{-8}$ s$^{-1}$. We also extend the same analysis to the dissipative version of the CSL model, finding in such a case less stringent bounds. FINIRE CON DISSIPATIVE BOOST AND NON WHITE.

%%%%%%%%%%%%%%%%%%%%%%%%%%%%%%%%%%%%%%%%%%%%%%%%%%%%%%%%%%%%%%%%%%%%%
\section*{Acknowledgements}
%%%%%%%%%%%%%%%%%%%%%%%%%%%%%%%%%%%%%%%%%%%%%%%%%%%%%%%%%%%%%%%%%%%%%
MB, SD and AB wish to thank H. Ulbricht for several discussions on the topic. They acknowledge  financial support from the University of Trieste. In addition, all authors acknowledge  financial support from INFN. %We thank H. Ulbricht and A. Vinante  for many useful and enjoyable discussions on the topic.
\appendix
\section{dSCL evolution during the delta-kick}
\label{sec:freedCSL}
\begin{figure}[t]
\includegraphics[scale=0.75]{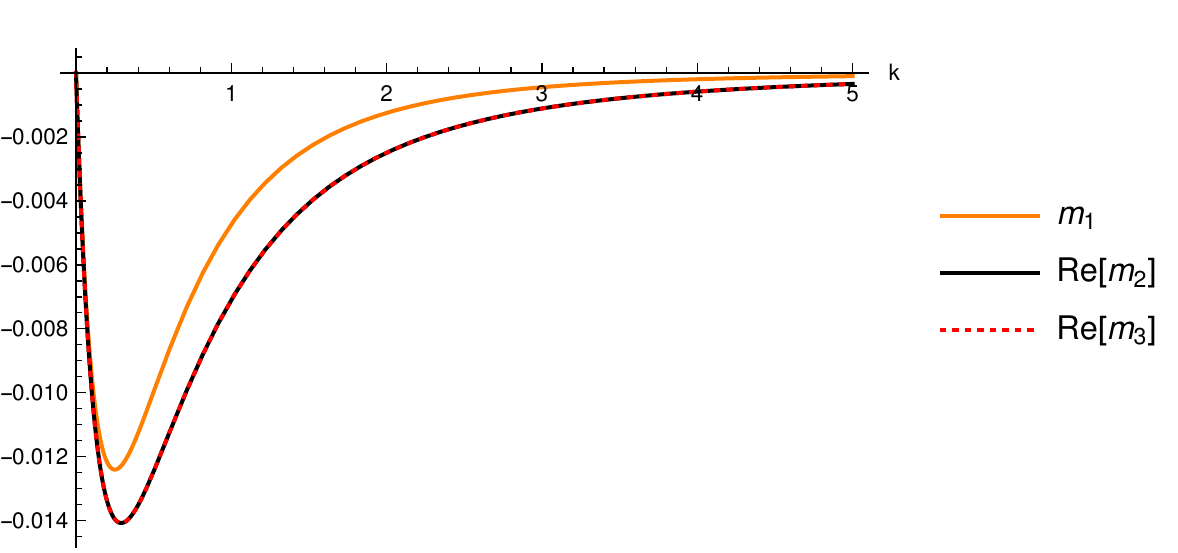}
\caption{Real part of the eigenvalues $m_i, \, i=1,2,3$ of matrix M in Eq.~\eqref{Hmatrix}, with $\lambda = 10^{-5} \text{s}^{-1}$, for different values of $k$.}
\label{autovalori1}
\end{figure}

\begin{figure}[t]
\includegraphics[scale=0.75]{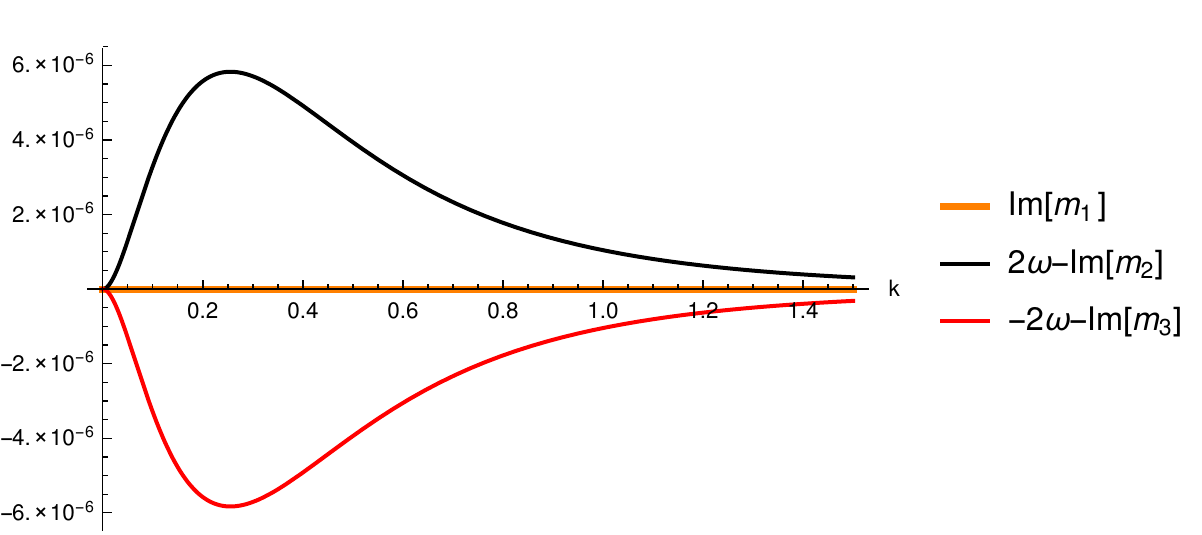}
\caption{Imaginary part of the eigenvalues $m_i, \, i=1,2,3$ of matrix M in Eq.~\eqref{Hmatrix}, with $\lambda = 10^{-5} \text{s}^{-1}$, for different values of $k$. Here $w=6,7$ rad/s.}
\label{autovalori2}
\end{figure}

We prove that during the delta-kick, the dCSL contribution to the dynamics is negligible with respect to the other effects. We start by rewriting the system of Eqs.~\eqref{x2dCSL3}, \eqref{xppxdcsl} and \eqref{dCSLp2} in terms of the dimensionless vector $\vec{x} (t) \in \mathbb{R}^3$ :
\begin{equation}
\label{defxbar}
\vec{x}_t := \begin{pmatrix}
\frac{m \omega}{\hbar} \left\langle \hat{\ve{x}}^{2}\right\rangle _{t} \\
\frac{1}{\hbar}\left\langle \hat{\ve{x}}\cdot\hat{\ve{p}}+\hat{\ve{p}}\cdot\hat{\ve{x}}\right\rangle _{t} \\
\frac{1}{\hbar m \omega}\left\langle \hat{\ve{p}}^{2}\right\rangle _{t}
\end{pmatrix} ,
\end{equation}
so that they take the form:
\begin{equation}
\label{sysdCSL2}
\frac{d}{dt}
\vec{x}_t
= 
\vec{f} + M \vec{x}_t,
\end{equation}
where 
\begin{equation}
\label{Hmatrix}
M = \begin{pmatrix}
0 & \omega & 0 \\
-2\omega & -B & 2 \omega \\
0& -\omega &-\chi
\end{pmatrix}
\end{equation}
and 
\begin{equation}
\label{fbar}
\vec{f} = \begin{pmatrix}
\frac{m \omega  \alpha}{\hbar} \\
0\\
\frac{ \chi\left\langle \hat{\ve{p}}^{2}\right\rangle _{as}}{\hbar m \omega}
\end{pmatrix}
.
\end{equation}
The formal solution of Eq.~\eqref{sysdCSL2} is given by
\begin{equation}
\label{harmsoldcsl}
\vec{x}_t = e^{M t} \vec{x}_0 + \int_{0}^{t} ds \, e^{M (t-s)} \vec{f}.
\end{equation}
We prove that the error done by neglecting the noise contributions ($\chi=B=\alpha=0$) is negligible, \textit{i.e.} that the exact solution Eq.~\eqref{harmsoldcsl} is well approximate by the quantum mechanical solution:
\begin{equation}
\label{qmsol}
\tilde{x}_t = e^{\tilde{M} t}  \vec{x}_0,
\end{equation}
where
\begin{equation}
\label{Htilde}
\tilde{M} =
\begin{pmatrix}
0 & \omega& 0 \\
-2\omega & 0 & 2\omega \\
0& -\omega &0
\end{pmatrix}
.
\end{equation}
While the quantum mechanical evolution \eqref{qmsol} is given by an unitary transformation \footnote{The eigenvalues of the matrix~\eqref{Htilde} are $0$, $\pm 2i\omega$.}, the dCSL dynamics involve a transient phase (until equilibrium is reached) expressed by the decaying exponential. In fact, the matrix \eqref{Hmatrix} has three distinct eigenvalues $m_1 \in \mathbb{R}$ and negative, and $m_2 = m_{3}^{*} \in \mathbb{C}$ with negative real parts, which are roots of the  characteristic third-order polynomial:
\begin{equation}
\label{charpoly}
m^3 + m^2 (B + \chi) + m (B\chi +4\omega^2) + 2\chi\omega^2.
\end{equation}
In Figs.~\ref{autovalori1} and~\ref{autovalori2}, the real parts of the eigenvalues $m_i, \, i = 1,2,3$ are shown as function of the dCSL parameter $k$ and with a relatively high collapse rate, $\lambda = 10^{-5} \text{s}^{-1}$. 

The quantum mechanical solution as given by Eq.~\eqref{qmsol} well approximates the exact dCSL solution as given by Eq.~\eqref{harmsoldcsl} if, for all  components $i = 1,2,3$,
\begin{equation}
\label{relation}
\frac{\abs{(\vec{x}_{t}^{i}-\vec{x}_{0}^{i}) - (\tilde{x}_{t}^{i} - \tilde{x}_{0}^{i})}}{\abs{\tilde{x}_{t}^{i} - \tilde{x}_{0}^{i}}} = \frac{\abs{\vec{x}_{t}^{i} - \tilde{x}_{t}^{i}}}{\abs{\tilde{x}_{t}^{i} - \tilde{x}_{0}^{i}}} \ll 1,
\end{equation}
where we used the equality $\vec{x}_{0}^{i}=\tilde{x}_{0}^{i}$. 

We start by noting that the numerator is limited by $\abs{\vec{x}_{t}^{i}- \tilde{x}_{t}^{i}} \leq \norma{\vec{x}_t - \tilde{x}_t}$, where $\norma{\cdot}$ is the usual Euclidean norm:
\begin{equation}
\label{eucl}
\norma{\vec{y}} = \sum_{i = 1}^{3} \abs{y_i}^2 .
\end{equation}
To proceed, it is convenient to use the matrix norm. From the Euclidean norm defined in Eq.~\eqref{eucl}, the following matrix norm can be defined (see~\cite{strocchi} for all the relevant properties of the matrix norm, as well as for the notation):
\begin{equation}
\label{matrnorm}
\norma{A} = \underset{\vec{x} \in \mathbb{R}^3}{ \text{sup}} \frac{\norma{A \vec{x}}}{\norma{\vec{x}}} = \underset{m \in \sigma(A^{\dagger}A)}{\text{max}} \abs{m},
\end{equation}
where $\sigma (B)$ is the \textit{spectrum} of the matrix $B$.
In particular, we will need the following  properties of the Euclidean  matrix norms:
\begin{align}
&\norma{A \vec{x}} \leq \norma{A} \norma{\vec{x}}, \label{sub1}\\
&\norma{AB} \leq \norma{A} \norma{B}, \label{sub2}
\end{align}
together with the triangular inequality
\begin{equation}
\label{triang}
\norma{A + B} \leq \norma{A} + \norma{B}.
\end{equation}
Taking into account the unitarity of the quantum evolution, and the exponential decay induced by dCSL, the following relation holds
\begin{equation}
\label{rel1}
\norma{e^{M t}} \leq \norma{e^{\tilde{M}t}} \leq 2.
\end{equation}
\begin{figure}[t!]
\includegraphics[scale=0.75]{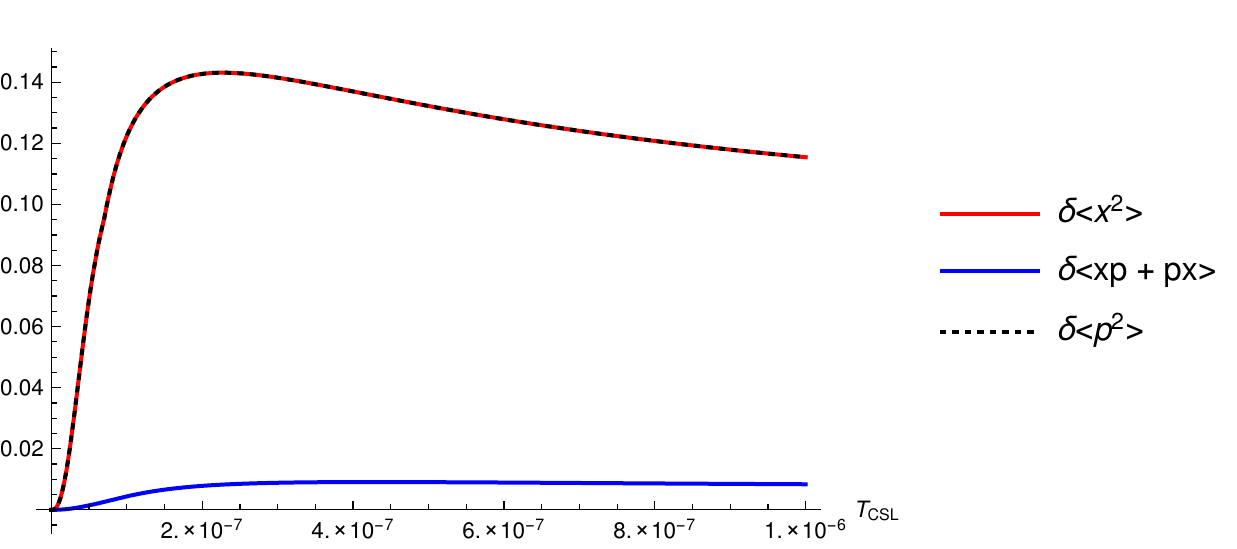}
\caption{Error functions defined in Eq.~\eqref{deferr} as functions of the CSL temperature $T_{\text{\tiny{CSL}}}$. Here we used $\lambda = 10^{-5} \text{ s}^{-1}$, $r_C = 10^{-7}$ m, $\delta t_2 = 35$ ms.}
\label{deltax-temp}
\end{figure}
\begin{figure}[t]
\includegraphics[scale=0.75]{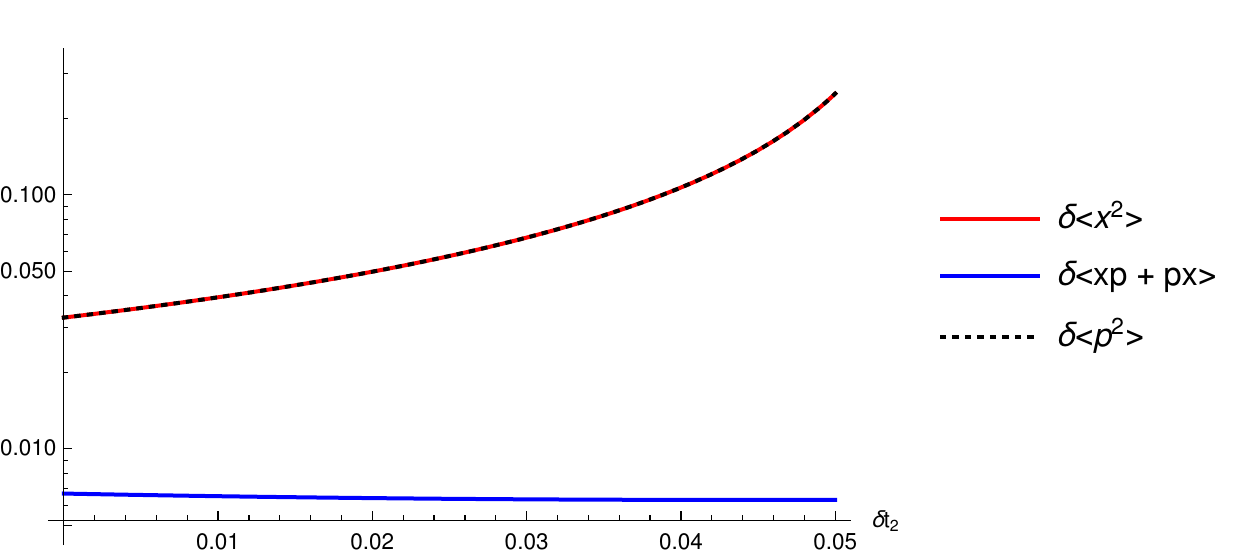}
\caption{ Log-plot of the error functions defined in Eq.~\eqref{deferr} as functions of the delta-kick time $\delta t_2$. Here we used $\lambda = 10^{-5} \text{ s}^{-1}$, $r_C = 10^{-7}$ m, $T_{\text{\tiny{CSL}}} = 1$ K.}
\label{deltax-taup}
\end{figure}
The last inequality in Eq.~\eqref{rel1} is obtained through a direct computation of the matrix norm defined in Eq.~\eqref{matrnorm}.
Using Eqs.~\eqref{harmsoldcsl} and \eqref{qmsol} together with the triangular inequality in Eq.~\eqref{triang}, we get:
\begin{equation}
\label{step1}
\begin{split}
&\norma{\vec{x}_t -\tilde{x}_t} = \norma{\left(e^{M t} - e^{\tilde{M}t}\right) \vec{x}_0 +  \int_{0}^{t} ds \, e^{M (t-s)} \vec{f}} \\
& \leq \norma{\left(e^{M t} - e^{\tilde{M}t}\right) \vec{x}_0} + \norma{ \int_{0}^{t} ds \, e^{M (t-s)} \vec{f}}.
\end{split}
\end{equation} 
Let us focus on the first term on the right hand side of Eq.~\eqref{step1}. Using the following identity,
\begin{equation}
\label{identity}
\begin{split}
&e^{M t} - e^{\tilde{M}t} = \int_{0}^{t} ds \, \frac{d}{ds} \left( e^{M s}e^{\tilde{M} (t-s)}\right) \\
&= \int_{0}^{t} ds \, e^{M s} \left(M - \tilde{M} \right) e^{\tilde{M} (t-s)}, 
\end{split}
\end{equation}
we can write:
\begin{equation}
\label{step2}
\begin{split}
&\norma{\left(e^{M t} - e^{\tilde{M}t}\right) \vec{x}_0} \leq \norma{\vec{x}_0} \times \\
& \times \int_{0}^{t} ds \, \norma{e^{M s} \left(M - \tilde{M} \right) e^{\tilde{M} (t-s)}} \\
& \leq \norma{\vec{x}_0} \int_{0}^{t} ds \, \norma{e^{Ms} \left(M - \tilde{M} \right)} \\
& \leq  \norma{\vec{x}_0} \,\norma{M - \tilde{M}} \int_{0}^{t} ds \, \norma{e^{Ms}} \\
& \leq  \norma{\vec{x}_0}\, \text{max} (B,\chi) 2 t.
\end{split}
\end{equation}
where in the first line we used Eq.~\eqref{sub1}, in the second and the third line Eq.~\eqref{sub2} together with $\norma{e^{\tilde{M}t}} \leq 2$, and in the last line Eq.~\eqref{rel1} and the matrix norm definition in Eq.~\eqref{matrnorm} for the diagonal matrix $M - \tilde{M}$.    

We now consider the second term in the second line of Eq.~\eqref{step1}. With a similar calculation as that in Eq.~\eqref{step2}, the following relation is found:
\begin{equation}
\label{step3}
\norma{\int_{0}^{t} ds \, e^{H (t-s)} \vec{f}} \leq 2 t \norma{\vec{f}}.
\end{equation}

Then, using the inequalities in eqs.~\eqref{step1}, \eqref{step2} and \eqref{step3}, the following upper bound on the error functions in Eq.~\eqref{relation} is found 
\begin{equation}
\label{finalrelation}
%\begin{split}
\frac{\abs{\vec{x}_{t}^{i} - \tilde{x}_{t}^{i}}}{\abs{\tilde{x}_{t}^{i} - \tilde{x}_{0}^{i}}} \leq \frac{2 t\left( \norma{\vec{x}_0} \text{max} (B,\chi) + \norma{\vec{f}} \right)}{\abs{\tilde{x}_{t}^{i}-\tilde{x}_{0}^{i}}} 
%\end{split}
\end{equation}
With reference to eq.~\eqref{defxbar}, we define the error bounds found in eq.~\eqref{finalrelation} as follows:
\begin{equation}
\label{deferr}
\frac{2 t\left( \norma{\vec{x}_0} \text{max} (B,\chi) + \norma{\vec{f}} \right)}{\abs{\tilde{x}_{t}^{i}-\tilde{x}_{0}^{i}}} =
\begin{cases}
 \! \delta \left \langle \hat{\ve{x}}^2 \right \rangle_t, \quad & i = 1; \\
 \! \delta \left \langle \hat{\ve{x}}\cdot\hat{\ve{p}}+\hat{\ve{p}}\cdot\hat{\ve{x}} \right \rangle_t, \quad &i=2; \\
 \! \delta \left \langle \hat{\ve{p}}^2\right \rangle_t, \quad &i = 3.
\end{cases}
\end{equation}
The values of the error functions defined in Eq.~\eqref{deferr} as a function of the delta-kick time $\delta t_2$ and the noise temperature $T_{\text{\tiny{CSL}}}$ are showed respectively in fig.~\ref{deltax-taup} and fig.~\ref{deltax-temp}. Despite having taken the strongest possible value for the collapse rate $\lambda = 10^{-5} \text{ s}^{-1}$ (as discussed in the introduction, larger values are  excluded by other experiments), the relative error is always below $0.14$, confirming the fact that the noise effects can be neglected during the delta-kick.

\end{document}